\documentclass[aps,twocolumn, pra, superscriptaddress, amsmath, amssymb, floatfix]{revtex4-1}
\usepackage{graphicx}        
\usepackage{amssymb}   
\usepackage{amsmath}
\usepackage{color,soul}
\usepackage{dcolumn}
\usepackage{bm}
\usepackage{mathtools}
\usepackage[table]{xcolor}
\usepackage{epstopdf}
\usepackage{float}
\usepackage{dsfont}

\newcommand{\ket}[1]{{\vert}#1{\rangle}}
\newcommand{\bra}[1]{{\langle}#1{\vert}}
\newcommand{\e}{\mathrm{e}}

\begin{document}
\title{Photonic dephasing dynamics and role of initial correlations}
%%%%%%%%%%%%%%%%%%%%%%%%%%%%%%%%%%%%%%%%%%%%
\date{\today}
%%%%%%%%%%%%%%%%%%%%%%%%%%%%%%%%%%%%%%%%%%%%
\author{Sina Hamedani Raja}
\affiliation{Turku Centre for Quantum Physics, Department of Physics and Astronomy,
University of Turku, FI-20014 Turun yliopisto, Finland}

\author{K P Athulya}
\affiliation{School of Physics,  Indian Institute of Science Education and Research, Thiruvananthapuram, Kerala, India 695551}

\author{Anil Shaji}
\affiliation{School of Physics,  Indian Institute of Science Education and Research, Thiruvananthapuram, Kerala, India 695551}

\author{Jyrki Piilo}
\affiliation{Turku Centre for Quantum Physics, Department of Physics and Astronomy,
University of Turku, FI-20014 Turun yliopisto, Finland}
%%%%%%%%%%%%%%%%%%%%%%%%%%%%%%%%%%%%%%%%%%%%
\begin{abstract}

Dynamics of open quantum systems depends on different types of initial correlations. On the one hand, 
 when system and environment are both inherently multipartite, initial correlations between the parties of the composite  environment make the dynamical map non-local, despite of local nature of the interaction between each party of the system and the environment. 
 On the other hand, initial correlations between the open system and its environment prevents one from defining a completely positive dynamical map. 
Recently, dephasing dynamics of photons has been used in both of these frameworks - theoretically and experimentally -  to demonstrate some fundamental and applicable aspects of open system dynamics and memory effects. However, the earlier studies in this context are often based solely on the concept of decoherence functions.
Therefore, we still lack explicit master equation descriptions for dynamics induced by correlated composite initial environmental states. Also, a detailed understanding how initial system-environment correlations influence qubit dynamics in the photonic context is missing.
In this paper, we derive generic master equations for the reduced dephasing dynamics of the two-photon polarization state when the bipartite environmental frequency degrees of freedom are initially correlated. We thereby show the explicit dependence of the operator form and the decay rates of the master equation on the initial frequency correlations and the influence of various types of frequency distributions. 
Furthermore, we use recently developed  \textit{bath positive decomposition} method to treat initially correlated polarization-frequency state of a photon, and demonstrate how this allows new insight and detailed information on how the contributions of different origin influence the photonic dephasing.
\end{abstract}
%%%%%%%%%%%%%%%%%%%%%%%%%%%%%%%%%%%%%%%%%%%%
\maketitle
\section{Introduction}

Understanding open system dynamics and decoherence is important in several areas of quantum physics~\cite{BreuerBook_2002,RivasBook_2012}. During the last ten years, there have been significant developments in both understanding the role of non-Markovian memory-effects~\cite{Rivas_2014, Breuer_2016, Hall_2018, Li_2019,Li_2019_EPL2} and in developing improved tools and techniques to treat open system 
dynamics~\cite{Vega_2017}.
Here, one of the common themes is the role that various types of correlations play in open system dynamics. In particular, understanding
 initial correlations between composite environments~\cite{Laine_2012,Liu_2013,Laine_2014,Xiang_2014,Liu_2016,HamedaniRaja_2017} and the role of initial system-environment correlations~\cite{pechukas94,Alicki_1995,pechukas95,jordan04,shaji05a,Linta-Shaji,Wiseman_2019,Lyyra_2018,Alipour_2019}  have led to fundamental insights as well as practical knowledge regarding open systems.

Photons  provide a common and  highly controllable system where the influence of correlations can be studied both conceptually and practically~\cite{Laine_2012,Liu_2013,Laine_2014,Xiang_2014,Liu_2016,HamedaniRaja_2017,Lyyra_2018}. Here, the polarization state of the photon is the open system and its frequency is the environment. Polarization and  frequency are coupled via birefringence leading to dephasing of a polarization state of the photon(s). The control of initial frequency distribution allows for the engineering of the decoherence and it is also possible to exploit various correlations for single photon or composite two-photon systems~\cite{ElsiNature2011,ElsiPRL2012,Lyyra_2018}.

On the one hand, dephasing dynamics of photons has often been described using the concept of decoherence functions and subsequent family of completely positive (CP) dynamical maps, in the past. On the other hand, master equations are one of the most common tools to treat open system dynamics~\cite{BreuerBook_2002}. However, master equations have not been used extensively when considering multipartite photonic systems and dephasing.
We consider first a bipartite two-photon system where the initial system-enviroment state is factorized whilst there exist initial correlations between the environmental states. 
It has been shown earlier that this induces non-local memory effects in open system dynamics~\cite{ElsiPRL2012,Liu_2013}. However,  the role of these types of initial correlations and non-local memory effects have not been considered on the level of master equations before, to the best of our knowledge. We derive generic master equations which display explicitly the role of initial correlations both on the dephasing rates and on the operator form of the master equation.  This allows also to reveal how even quite straightforward changes in the initial environmental state change drastically the description of photonic dephasing and increases the number of jump operators in the master equation. 

Continuing within the framework of correlations and open systems, we also study another long-standing problem in this context.
This is the role that initial system-environment correlations play in open system dynamics. Here, our interest is to see, what kind of insight the recently developed
\textit{bath positive decomposition} method~\cite{Wiseman_2019} allows when studying the open dynamics of the polarization states. This very general method is based on decomposing initial arbitrary system-environment 
state to a number of terms where each term can be treated with its individual CP-map. We show that for single-photon dephasing, this decomposition allows to describe, in a insightful way, how initial correlations influence the dynamics beyond the contribution arising from the factorized part.

The structure of the paper is the following. In the next section we describe briefly the basics of photonic dephasing.
In Section \ref{sec3} we focus on the correlations within the composite environment and derive various master equations in this context and discuss the insight they provide.
Section \ref{sec4}, in turn, describes the initially correlated system-environment case for single photon and Sec.~\ref{sec5} concludes the paper.

%%%%%%%%%%%%%%%%%%%%%%%%%%%%%%%%%%%%%%%%%%%%%%%%%%%%%%%%%%%%%%%%%%%%%% 
\section{Preliminaries with single photon dephasing dynamics}

We start with a brief recall of the single-photon dephasing model~\cite{ElsiNature2011}.
Polarization degree of freedom and frequency degree of freedom of a photon correspond to the open system and its environment, respectively. To begin with, we consider initially factorized joint polarization-frequency state 
\begin{equation}
\hat{\rho}_{SE}(0)=\hat{\rho}_S(0)\otimes \ket{\Omega}\bra{\Omega}.
\end{equation}
Here,  $\hat{\rho}_S(0)$ is the density operator of the initial polarization state and 
\begin{equation}
\ket{\Omega}= \int d \omega \; g(\omega)\ket{\omega},
\end{equation}
is the initial  frequency state where $g(\omega)$ is the probability amplitude that the photon has frequency $\omega$. 
The polarization Hilbert space is discrete and spanned by the horizontal-vertical polarization basis $\{\ket{h},\ket{v}\}$, while the Hilbert space of the frequency degree of freedom is spanned by the continuous frequency basis $\{\ket{\omega}\}$. 

The system-environment -- or polarization-frequency -- interaction is provided by the Hamiltonian ($\hbar=1$)
\begin{equation}\label{eq:Hamiltonian}
\hat{H}=(n_h\ket{h}\bra{h}+n_v\ket{v}\bra{v})\otimes\int d \omega \; \omega \; \ket{\omega}\bra{\omega} ,
\end{equation}
where $n_h$ ($n_v$) is the refraction index for polarization component $h$ ($v$).
For interaction time $t$, and tracing over the frequency, the reduced polarization state is
\begin{equation}\label{eq:MapSinglePhoton}
\hat{\rho}_S(t)=\begin{pmatrix}
\bra{ h} \rho_S(0)\ket{h} & \kappa(t)\bra{ h} \rho_S(0)\ket{v} \\
\kappa(t)^{*}\bra{ v} \rho_S(0)\ket{h} & \bra{ v} \rho_S(0)\ket{v}.
\end{pmatrix},
\end{equation}
Here, the dephasing dynamics is given by the decoherence function
\begin{equation}
\label{eq:kappa}
 \kappa(t)=\int d\omega \; \vert g(\omega)\vert^2 \mathrm{e}^{-i \Delta n \omega t},
 \end{equation}
where $\Delta n \equiv n_v-n_h$. Note that $0\leq \vert \kappa(t) \vert \leq 1 $ for all times $t\geq 0$ and $\vert \kappa(0) \vert=1$. 

Equation \eqref{eq:MapSinglePhoton} describes a $t$-parametrized completely positive (CP) map $\hat{\Phi}_t$, such that $\hat{\rho}_S(t)=\hat{\Phi}_t[\hat{\rho}_S(0)]$, and its corresponding master equation takes the form
 \begin{equation}\label{eq:MasterEq}
\frac{d}{dt}\hat{\rho}_S(t)=-i\frac{\nu(t)}{2}[\hat{\sigma}_z,\hat{\rho}_S(t)]+\frac{\gamma(t)}{2}[\hat{\sigma}_z \hat{\rho}_S(t) \hat{\sigma}_z-\hat{\rho}_S(t)].
\end{equation}
Here,  $\hat{\sigma}_z$ is the Pauli $z$ operator and the rates $\nu(t)$ and $\gamma(t)$ can be expressed in terms of the decoherence function $\kappa(t)$ as
\begin{equation}\label{eq:Rates}
\gamma(t)=-\Re \bigg[\frac{1}{\kappa(t)} \frac{d\kappa(t)}{dt}\bigg], \quad \nu(t)=-\Im \bigg[\frac{1}{\kappa(t)} \frac{d\kappa(t)}{dt}\bigg],
\end{equation}
where, $\Re[\cdot]$ and $\Im[\cdot]$ indicate the real and imaginary parts, respectively. 

Equation~\eqref{eq:Rates} shows that once the decoherence function $\kappa(t)$ is obtained from Eq.~\eqref{eq:kappa}, then we can derive the corresponding rates in master equation ~\eqref{eq:MasterEq}. Indeed,  the decoherence function $\kappa(t)$ in Eq.~\eqref{eq:kappa} is the Fourier transformation of the initial frequency 
probability distribution $P(\omega) =|g(\omega)|^2$, and therefore the control of this distribution allows to study various types of dephasing maps and to engineer  the form and time dependence of the dephasing rate $\gamma(t)$ in master equation ~\eqref{eq:MasterEq}.

For example, a Gaussian frequency distribution with variance $\sigma^2$ and mean value $\bar{\omega}$, i.e., 
\[P(\omega)=\frac{\mathrm{exp}[-(\omega-\bar{\omega})^2/2\sigma^2]}{\sqrt{2\pi}\sigma},\]
leads to a positive and time dependent dephasing rate $\gamma(t)= \Delta n ^2 \sigma^2 t$ which presents a time-dependent Markovian dynamics. On the other hand, a Lorentzian distribution 
\[P(\omega)=\frac{\lambda}{\pi[(\omega-\omega_0)^2+\lambda^2]},\]
results in a constant decay rate $\gamma=\lambda \Delta n $, corresponding to dynamical semi-group and Lindbad-Gorini-Kossakowski-Sudarshan (LGKS) dynamics~\cite{GoriniJmathPhys1976,LindbladCommun1976}. We note that the latter case has been also reported in \cite{Smirne2014, Smirne2018}. 
The transition from Markovian to non-Markovian regime, in turn,  is observed with further modifications of the frequency distribution~\cite{ElsiNature2011}.

In the following, we generalize the master equation~in Eq.\eqref{eq:MasterEq} to two-photon case. In particular, we are interested in how the initial correlations between the frequencies of the two photons influence the various dephasing rates and the operator form of the corresponding master equation for a bipartite open system. 
%%%%%%%%%%%%%%%%%%%%%%%%%%%%%

\section{Master equation for two-photon dephasing dynamics: role of initially correlated joint frequency distribution} \label{sec3}

%\subsection{Two-photon dephasing dynamics: derivation of the master equation in presence of frequency correlation} \label{secIIb}
Consider a pair of photons, labeled $a$ and $b$, whose total polarization-frequency initial state is again in a factorized form
\begin{equation}
\hat{\rho}_{SE}(0)=\hat{\rho}_S(0)\otimes \ket{\Omega}\bra{\Omega},
\end{equation}
where now
\begin{equation}
\ket{\Omega}=\int d \omega_a \int  d \omega_b \; g(\omega_a,\omega_b)\ket{\omega_a,\omega_b},
\end{equation}
is the initial state of the two-photon frequency degree of freedom and the corresponding joint probability distribution is $P(\omega_a,\omega_b)=\vert g(\omega_a,\omega_b)\vert ^2$. Initial polarization state is $\hat{\rho}_S(0)$, whose Hilbert space is spanned by the bipartite basis $\{\ket{hh},\ket{hv},\ket{vh},\ket{vv}\}$. 

The polarization of each photon interacts locally with its own frequency and therefore system-environment interaction Hamiltonian for the two photons is a sum of the two local contributions~\cite{ElsiPRL2012}
\begin{equation}\label{eq:HamiltonianTwoPhoton}
\hat{H}=\hat{H}_a \otimes \hat{I}_b +  \hat{I}_a \otimes \hat{H}_b.
\end{equation}
Here, each local term is given by Eq.~\eqref{eq:Hamiltonian} and $\hat{I}_a$ ($\hat{I}_b$) is the identity operator for photon a (b).

We write initial bipartite polarization state $\hat{\rho}_S(0)$ as
\[\hat{\rho}_S(0)=\sum_{\alpha,\beta}\sum_{\alpha',\beta'}p_{\alpha\beta,\alpha'\beta'}\ket{\alpha \beta}\bra{\alpha' \beta'},\]
with sums over $h$ and $v$. After interaction time $t$, the polarization state is \citep{ElsiPRL2012}
\begin{flalign}\label{eq:MapTwoPhotoon}
&\hat{\rho}_S(t)=
\\ 
&\begin{pmatrix} \nonumber
p_{hh,hh} & \kappa_b(t)p_{hh,hv} & \kappa_a(t)p_{hh,vh} & \kappa_{ab}(t)p_{hh,vv} \\
\kappa_b^{*}(t) p_{hv,hh} & p_{hv,hv} & \Lambda_{ab}(t)p_{hv,vh} & \kappa_a(t)p_{hv,vv} \\
\kappa_a^{*}(t) p_{vh,hh} & \Lambda_{ab}^{*}(t)p_{vh,hv} & p_{vh,vh} & \kappa_b(t)p_{vh,vv} \\
\kappa_{ab}^{*}(t) p_{vv,hh} & \kappa_a^{*}(t)p_{vv,hv} & \kappa_b(t)^{*} p_{vv,vh} & p_{vv,vv}
\end{pmatrix}.
\end{flalign}
Here, the local decoherence functions for photon $j=a,b$ are given by
\begin{equation}
\label{k-loc}
\kappa_j(t)=\int d\omega_a  \int \;d\omega_b \; \vert g(\omega_a, \omega_b)\vert^2 \mathrm{e}^{-i \Delta n \omega_j t},
\end{equation}
and the non-local ones by
\begin{equation}
\label{k1-nloc}
\kappa_{ab}(t)=\int d\omega_a \int \; d\omega_b \; \vert g(\omega_a, \omega_b)\vert^2 \mathrm{e}^{-i \Delta n(\omega_a+\omega_b)t},
\end{equation}
and 
\begin{equation}
\label{k2-nloc}
\Lambda_{ab}(t)=\int d\omega_a \int \; d\omega_b \;\vert g(\omega_a, \omega_b)\vert^2 \mathrm{e}^{-i \Delta n(\omega_a-\omega_b)t}.
\end{equation}
The density matrix evolution given by Eqs.~(\ref{eq:MapTwoPhotoon}-\ref{k2-nloc})
can also be described by a $t$-parametrized completely positive dynamical map $\hat{\Phi}_t$, such that 
\begin{equation}\label{eq:DynMap}
\hat{\rho}_S(t)=\hat{\Phi}_t[\hat{\rho}_S(0)].
\end{equation} It is important to note that when the initial joint frequency distribution factorizes,  $P(\omega_a,\omega_b)=P_a(\omega_a)\times P_b(\omega_b)$, then the global decoherence functions are products of the local ones, i.e.,
$\kappa_{ab} (t) = \kappa_a (t)  \kappa_b (t)$ and $\Lambda_{ab} (t)=  \kappa_a (t)  \kappa^{\ast}_b (t)$. Subsequently, the map for the bipartite photon system is tensor product of the local CP maps $\hat{\Phi}_t=\hat{\Phi}_t^{(a)} \otimes \hat{\Phi}_t^{(b)}$. However, when the initial frequency distribution does not factorize, $P(\omega_a,\omega_b)\neq P_a(\omega_a)\times P_b(\omega_b)$, and contains correlations, then the map for the bipartite system is not anymore product of the local maps, $\hat{\Phi}_t\neq
\hat{\Phi}_t^{(a)} \otimes \hat{\Phi}_t^{(b)}$~\cite{ElsiPRL2012}. Now, we are interested in how to derive the generator of the corresponding non-local bipartite dynamical map and  what are the modifications in the corresponding  dephasing master equations when the amount of initial frequency correlations change.

We begin our derivation by writing the dynamical map formally as
\begin{equation}\label{eq:dynMap}
\hat{\Phi}_t = \exp \Big[\int_0^t d\tau \hat{\mathcal{L}}_{\tau}\Big],
\end{equation}
where $\hat{\mathcal{L}}_t$ is the generator of the dynamics. Finding an expression for the generator then provides us the master equation we want to construct as
\begin{equation}\label{eq:formalME}
\frac{d}{dt}\hat{\rho}_S(t)=\hat{\mathcal{L}}_t [\hat{\rho_S}(t)].
\end{equation}
Provided that the map in Eq.~\eqref{eq:dynMap} is invertible and its derivative is well-defined, one can obtain the generator as
\begin{equation}\label{eq:genMapInvMap}
\mathcal{\hat{L}}_t=\frac{d}{dt}\hat{\Phi}_t \circ \hat{\Phi}^{-1}_t.
\end{equation}
%Note that the dynamical map may not be invertible in general, however, let us assume this requirement is fulfilled for now and postpone a detailed discussion to a later section. 
To find the generator in Eq.~\eqref{eq:genMapInvMap} we need a suitable representation for the dynamical map $\hat{\Phi}_t$. With this in mind, we expand the two-photon density matrix $\hat{\rho}_S(t)$ in terms of a complete and orthonormal operator basis $\{\hat{F_{\alpha}}\}$. Specifically, we choose here fifteen generators of $\mathrm{SU}(4)$, whose exact  expressions can be found in~\cite{Alicki1987}, plus $\hat{F}_1=\hat{I}/\sqrt{4}$, such that $\mathrm{Tr}[\hat{F}_i^{\dagger} \hat{F}_j]=\delta_{i j}$. It is worth mentioning that one can alternatively use the basis constructed by the tensor product of Pauli matrices plus the identity.  Fixing the basis for the representation, then the two-photon polarization state at time $t$ is
\begin{equation}\label{eq:BlochVec1}
\hat{\rho}_S(t)=\sum_{\alpha=1}^{16} r_{\alpha} (t)\hat{F}_{\alpha}, \qquad r_{\alpha}(t)= \mathrm{Tr}[\hat{F}_{\alpha}\hat{\rho}_S(t)],
\end{equation}
where coefficients $\{r_{\alpha}\}$ form the generalized Bloch vector corresponding to the state $\hat{\rho}_S(t)$ as
\begin{equation}\label{eq:BlochVec2}
\vec{r}(t)=(1/2,r_2(t),...,r_{16}(t))^{\mathrm{T}}.
\end{equation} 
By using Eq.~\eqref{eq:BlochVec1} for both $\hat{\rho}_S(t)$ and $\hat{\rho}_S(0)$, we can write Eq.~\eqref{eq:DynMap} as
\begin{equation}\label{eq:MapBlochRep}
r_{\alpha}(t)=\sum_{\beta}[\hat{\Phi}_t]_{\alpha \beta} r_{\beta}(0),
\end{equation}
where $[\hat{\Phi}_t]$ is the transformation matrix corresponding to the map $\hat{\Phi}_t$ represented in the basis $\{\hat{F_{\alpha}}\}$. Elements of this matrix depend on the decoherence functions given in 
Eqs.~(\ref{k-loc}-\ref{k2-nloc}) and each column can be systematically calculated by using a proper pair of initial and evolved states (c.f.~Eq.~\eqref{eq:MapTwoPhotoon}). One can proceed to find the matrix representation of the generator by calculating the derivative and inverse of $[\hat{\Phi}_t]$ and using them in Eq.\eqref{eq:genMapInvMap}, such that
\begin{equation}\label{eq:genMat}
[\hat{\mathcal{L}}_t]=\frac{d}{dt}[\hat{\Phi}_t].[\hat{\Phi}_t]^{-1},
\end{equation}
where we have replaced operator multiplication by matrix multiplication.

Let us now consider the generator in a Lindblad operator form
\begin{eqnarray}\label{eq:LindbladGen}
\hat{\mathcal{L}}_t[\hat{\rho}_S(t)]&&=-i[\hat{H}(t),\hat{\rho}_S(t)]+
\\
&&\sum_{\alpha=2}^{16}\sum_{\beta=2}^{16} R_{\alpha \beta}(t)\Big(\hat{F}_{\alpha}\hat{\rho}_S(t) \hat{F}_{\beta}^{\dagger}-\frac{1}{2}\{\hat{F}_{\beta}^{\dagger}\hat{F}_{\alpha},\hat{\rho}_S(t)\}\Big), \nonumber
\end{eqnarray}
where 
\begin{equation}
\hat{H}(t)=\frac{-1}{2i}\sum_{\alpha=2}^{16}\Big[R_{\alpha1}(t)\hat{F}_{\alpha}-R_{1\alpha}(t)^{*}\hat{F}^{\dagger}_{\alpha}\Big],
\end{equation}
captures the environment induced coherent dynamics and $R_{\alpha \beta}(t)$ with $\alpha, \beta =2,3,...,16$ are elements of a $15\times 15$ matrix providing the decay rates. Each element in the matrix representation of the generator then reads
\begin{equation}\label{eq:equality}
[\hat{\mathcal{L}}_t]_{\alpha\beta}=\mathrm{Tr}[\hat{F}_{\alpha}^{\dagger} \hat{\mathcal{L}}_t[\hat{F}_{\beta}]].
\end{equation}
Here we use Eq.~\eqref{eq:LindbladGen} in the right hand side. Finally, by elementwise comparison of Eq.~\eqref{eq:equality} with Eq.~\eqref{eq:genMat} we find the decay rates of the Lindblad master equation in Eq.~\eqref{eq:LindbladGen} in terms of the decoherence functions in Eqs.~(\ref{k-loc}-\ref{k2-nloc}).
Before proceeding further, let us note that generator of a CP-divisible map always has a Linblad form \cite{Alicki1987,GoriniJmathPhys1976,LindbladCommun1976,RivasBook_2012}. A map $\hat{\Phi}_t$ is CP-divisible  if it can be decomposed as $\hat{\Phi}_t=\hat{\Phi}_{t,s}\hat{\Phi}_s$ where the intermediate map $\hat{\Phi}_{t,s}$ is also a legitimate CP-map for all $t \geqslant s \geqslant0 $ \cite{RivasPRL2010}. In this paper, however, we do not restrict ourselves to the CP-divisible maps and as we show later we also take non-Markovian dynamics into account.

After finding the general expression for the decay rate matrix, it turns out that it is quite sparse and can be reduced to a $3\times3$ matrix, which we denote by $R(t)$. The corresponding subspace is spanned by only three generators of $\mathrm{SU}(4)$, which are linearly dependent on the operators $\hat{I}_2 \otimes \hat{\sigma}_z$, $\hat{\sigma}_z \otimes \hat{I}_2$, and $\hat{\sigma}_z \otimes \hat{\sigma}_z$. This is indeed intuitive because population elements of the density matrix are invariant upon a dephasing channel, so those terms that couple the levels must be absent. 
The explicit expression for the matrix $R(t)$, corresponding to a general frequency distribution, is provided in the Appendix. Considering this general result, we diagonalize it to rewrite the second term on r.h.s of Eq.~\eqref{eq:LindbladGen} in the form
\begin{equation}\label{eq:NormalizedME}
\hat{\mathcal{D}}[\hat{\rho}_S(t)]=\sum_{\alpha=1}^{3} \gamma_{\alpha}(t)\Big[\hat{J}_{\alpha}\hat{\rho}_S(t)\hat{J}_{\alpha}^{\dagger}-\frac{1}{2}\{\hat{J}_{\alpha}^{\dagger}\hat{J}_{\alpha},\hat{\rho}_S(t)\}\Big],
\end{equation}
where 
\begin{equation}
\begin{pmatrix}
\gamma_1(t) & 0 & 0 \\
0 & \gamma_2(t) & 0 \\
0 & 0 & \gamma_3(t)
\end{pmatrix}=U R(t) U^{\dagger}, \quad \quad \hat{J}_{\alpha}=\sum_j U_{\alpha j}\hat{F}_j,
\end{equation}
and $U$ is the orthogonal transformation which diagonalizes the matrix $R(t)$. It is worth stressing that if the dynamical map in hand is CP-divisible, then all decay rates will be non-negative, i.e. $\gamma_i(t)\geq 0$ for all interaction times $t\geq 0$. 

Above general results hold for arbitrary initial frequency distributions.
In the following, we discuss explicitly initially correlated joint frequency distributions for bivariate single- and double-peak Gaussian cases.
These choices are motivated by their use in recent theoretical and experimental works, see e.g.~\cite{ElsiNature2011,ElsiPRL2012,HamedaniRaja_2017}, and due to their ability to account for the explicit influence of frequency correlations in the dephasing dynamics.

%%%%%%%%%%%%%%%%%%%%%%%%%%%%% 
\subsection{Single-peak bivariate Gaussian distribution}

Consider the joint bivariate  Gaussian frequency distribution $P_{ab}(\omega_a, \omega_b)$
and its covariance matrix $C$, such that $C_{ij}=\langle \omega_i \omega_j\rangle-\langle\omega_i\rangle\langle \omega_j \rangle$ for $i,j=a,b$ \citep{ElsiPRL2012}.
The correlation coefficient is now given by $ K=C_{ab}/\sqrt{C_{aa}C_{bb}}$, such that $-1\leq K \leq 1$. A fully anti-correlated initial frequency distribution has $K=-1$, which dictates that for any pair of $\omega_a$ and $\omega_b$ we have $\omega_a+\omega_b\equiv \omega_0$, with some constant frequency $\omega_0$.
The means of the local single photon frequency distributions are given by $(\bar{\omega}_a,\bar{\omega}_b)^{\mathrm{T}}$ and we denote the difference between the local means as $\bar{\omega}_a-\bar{\omega}_b =\Delta\omega$ and their sum as $\bar{\omega}_a+\bar{\omega}_b =\omega_0$. 
Using Eqs.~(\ref{k-loc}-\ref{k2-nloc}) and denoting the variance of the distribution by $\sigma^2$,  the decoherence functions become
\begin{flalign}
&\kappa_a(t)=\mathrm{exp}\Big[\frac{-\sigma^2 \Delta n^2 t^2-i \Delta n t (\omega_0+\Delta\omega)}{2}\Big],
\\
&\kappa_b(t)=\mathrm{exp}\Big[\frac{-\sigma^2 \Delta n^2 t^2-i \Delta n t (\omega_0-\Delta\omega)}{2}\Big],
\\
&\kappa_{ab}(t)=\mathrm{exp}\big[-\sigma^2 \Delta n^2 t^2(1+K)-i \Delta n t \omega_0\big],
\\
&\Lambda_{ab}(t)=\mathrm{exp}\big[-\sigma^2 \Delta n^2 t^2(1-K)-i \Delta n t \Delta\omega\big].
\end{flalign}
It is straightforward to check that the corresponding transformation matrix $[\hat{\Phi}_t]$ for the generalized Bloch vector is always invertible when time $t$ is finite. 
After inserting the above expressions for the decay rate matrix $R(t)$, see the Appendix, and followed by diagonalization, we obtain the rates  appearing in the master equation~\eqref{eq:NormalizedME} as follows
\begin{flalign}
&\gamma_1(t)=2(1-K)\sigma^2 \Delta n^2 t,\label{eq:gamma1}
\\
&\gamma_2(t)=2(1+K)\sigma^2 \Delta n^2 t,\label{eq:gamma2}
\\
&\gamma_3(t)=0,\label{eq:gamma3}
\end{flalign}
and the corresponding jump operators 
\begin{flalign}
&\hat{J}_1=\frac{1}{2\sqrt{2}}(\hat{I}_2\otimes \hat{\sigma}_z + \hat{\sigma}_z \otimes \hat{I}_2),\label{eq:J1}
\\
&\hat{J}_2=\frac{1}{2\sqrt{2}}(\hat{I}_2\otimes \hat{\sigma}_z - \hat{\sigma}_z \otimes \hat{I}_2),\label{eq:J2}
\\
&\hat{J}_3=\frac{1}{2}\hat{\sigma}_z \otimes \hat{\sigma}_z.\label{eq:J3}
\end{flalign}

\begin{figure}[t]
\centering
\includegraphics[width=0.95 \columnwidth, height = 1.5 cm]{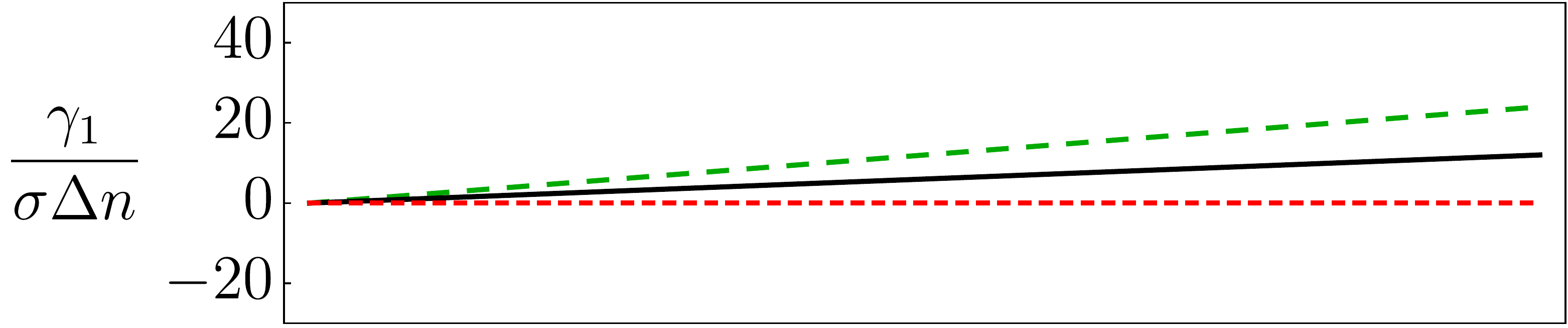}
\includegraphics[width=0.95 \columnwidth, height = 1.5 cm]{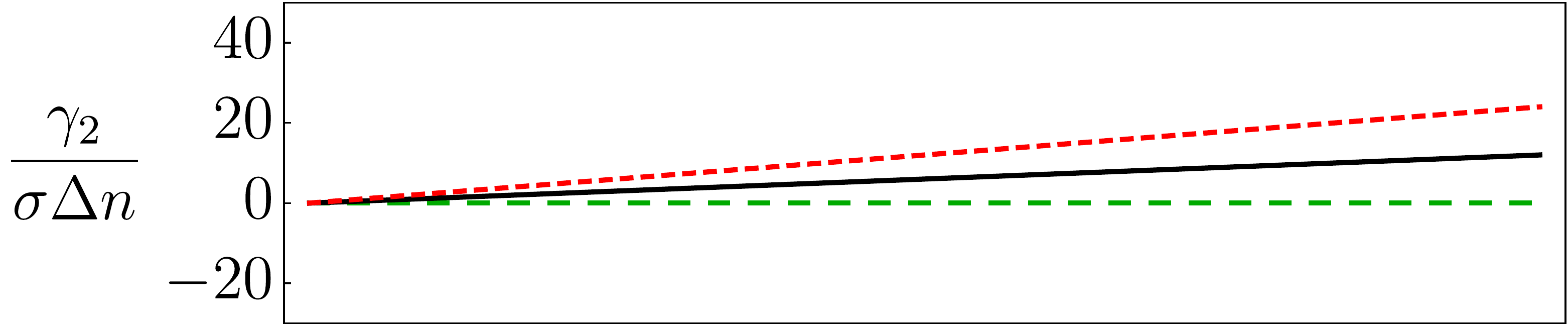}
\includegraphics[width=0.95 \columnwidth ,height = 2.4 cm]{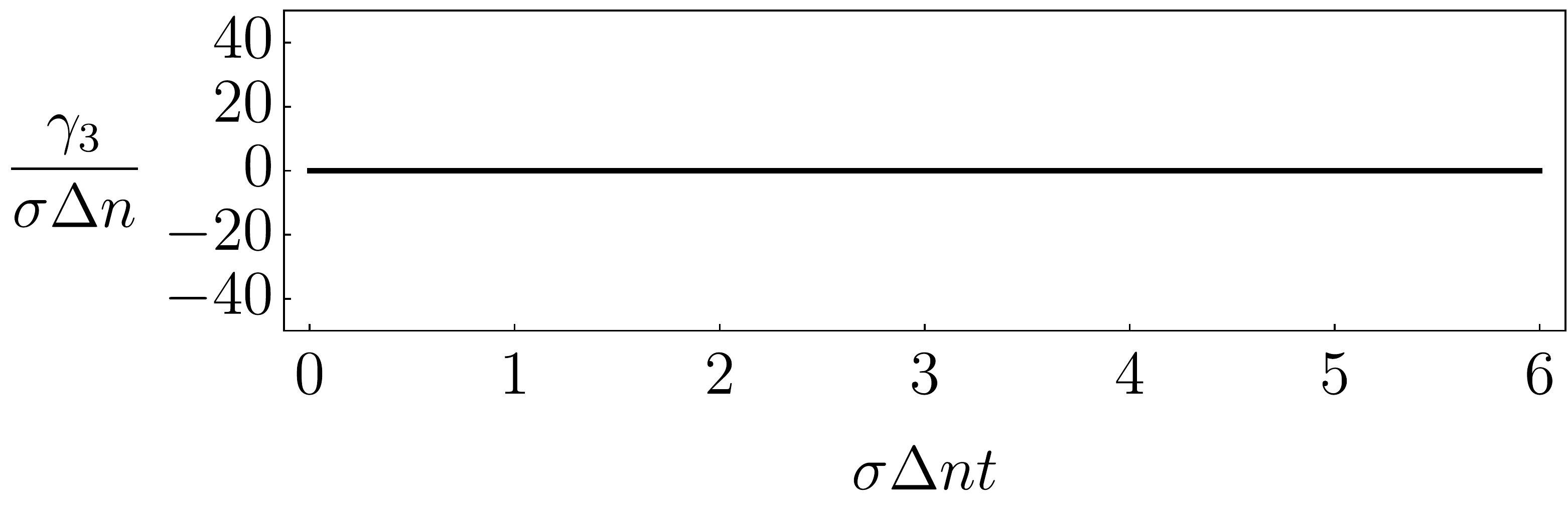}
\caption{(Color online) Decay rates as a function of normalized interaction time in the case of single-peak Gaussian frequency distribution. Large-dashed green when $K=-1$, solid black when $K=0$, and small-dashed red when $K=-1$. Here we set $\Delta \omega /\sigma=2$.}
\label{fig1}
\end{figure}

Dephasing rates $\gamma_1$ and $\gamma_2$ are linear functions of time and their slopes depend on the correlation coefficient $K$. Figure~\ref{fig1} displays the rates for  $K=-1,0,1$. Since all the rates are non-negative and the first two are time dependent, this leads to CP-divisible dynamics which, however, does not fulfil the LGKS semigroup property. It is also interesting to note here the absence of the jump operator $\hat{\sigma}_z \otimes \hat{\sigma}_z$ since the corresponding rate $\gamma_3$ is always equal to zero. 
Moreover, the role of the environmental correlation coefficient $K$ of the initial joint frequency distribution is now explicit in expressions~(\ref{eq:gamma1}-\ref{eq:gamma3}). When $K=1$ ($K=-1$) the rate $\gamma_1=0$ ($\gamma_2=0$) and we are left with only one dephasing channel given by $\hat{J}_2$ ($\hat{J}_1$).  When there are no initial correlations between the two environments, $K=0$, then 
$\gamma_1(t)=\gamma_2(t)$. Subsequently, the corresponding generator and master equation contain equally weighted contributions of 
 the two local jump operators $\hat{J}_1$ and $\hat{J}_2$. Changing the value of  the initial correlations $K$ allows then to tune the dynamics between the above mentioned extreme cases.

It is also worth discussing similarities and differences between our photonic model and the two-qubit model interacting with a common environment
 \cite{PalmaPROC1996, ReinaPRA2002, CironeNEWJPhys2009, AddisPRA2013}. In the latter model two qubits are spatially separated by  a distance $D$, while they both interact with same physical and common bosonic environment. It is interesting that the master equation describing this model has the exact same operator form and jump operators \cite{AddisPRA2013} obtained in equations \eqref{eq:J1} to \eqref{eq:J3}. In addition, decay rates derived in \cite{AddisPRA2013} exhibit similar dependence on the distance $D$, as our decay rates here depend on the correlation coefficient $K$. Moreover, when $D \rightarrow \infty$, the dynamical map will be factorized to $\hat{\Phi}_t=\hat{\Phi}_t^{(a)} \otimes  \hat{\Phi}_t^{(b)}$, with the superscripts corresponding to each qubit. The same behavior is also captured here when $K \rightarrow 0$.
 However, it is worth keeping in mind that in our case the two environments are distinct physical entities and the tuning of the generator -- or form of the master equation -- is obtained by changing the initial bipartite environmental state. Furthermore, we can tune the generator continuously between the fully correlated and anti-correlated cases.
 %%%%%%%%%%%%%%%%%%%%%%%%%%%%% 
%%%%%%%%%%%%%%%%%%%%%%%%%%%%%
%%%%%%%%%%%%%%%%%%%%%%%%%%%%%%%%%%%%%%%%%%%%%%%%%%%%%%%%%% 

\subsection{Double-peak bivariate Gaussian distribution}

We consider a double-peak frequency distribution as sum of two single-peak bivariate Gaussian distributions, already used in \cite{HamedaniRaja_2017}, such that 
\begin{equation}
P(\omega_a,\omega_b)=[P_1(\omega_a, \omega_b)+ P_2(\omega_a, \omega_b)]/2.
\end{equation}
We assume that both single-peak terms have the same correlation coefficient $K$ and standard deviation $\sigma$, but their means are located at $(\omega_0/2-\Delta \omega/2,\omega_0/2+\Delta \omega/2)^{\mathrm{T}}$ and $(\omega_0/2+\Delta \omega/2,\omega_0/2-\Delta \omega/2)^{\mathrm{T}}$, respectively. Please note that the correlation coefficient $K$ of each single-peak distribution $P_1(P_2)$ does not equal to the actual correlation coefficient of the bivariate distribution $P$, obtained by its covariance matrix. In more detail, whenever we have nonzero $K$ for each single-peak, we have non-zero correlation in $P$. But note that if $K=0$, then we still have correlation in $P$ as long as we have non-zero peak separation, $\Delta \omega \neq 0$.

The decoherence functions calculated from Eqs.~(\ref{k-loc}-\ref{k2-nloc}) become
\begin{flalign}
&\kappa_a(t)=\mathrm{exp}\bigg[\frac{-\sigma^2 \Delta n^2 t^2-i t \Delta n \omega_0}{2}\bigg]\cos\bigg(\frac{t \Delta n \Delta \omega}{2}\bigg),
\\
&\kappa_b(t)=\kappa_a(t),
\\
&\kappa_{ab}(t)=\mathrm{exp}\big[-\sigma^2 \Delta n^2 t^2 (1+K)-i t \Delta n \omega_0\big],
\\
&\Lambda_{ab}(t)=\mathrm{exp}\big[-\sigma^2 \Delta n^2 t^2 (1-K)\big]\cos(t \Delta n \Delta \omega).
\end{flalign}
By using the earlier obtained general results, in a similar manner compared to single-peak case, we obtain the dephasing
rates
\begin{flalign}
&\gamma_1(t)=2(1-K)\sigma^2 \Delta n^2 t+ \tan(t \Delta n  \Delta \omega )\Delta n \Delta \omega ,\label{eq:gamma1Two}
\\
&\gamma_2(t)=2(1+K)\sigma^2 \Delta n^2 t,\label{eq:gamma2Two}
\\
&\gamma_3(t)= \frac{1}{2} \tan \bigg(\frac{t \Delta n \Delta \omega}{2} \bigg)\big[1-\sec (t \Delta n \Delta \omega)\big]\Delta n \Delta \omega.\label{eq:gamma3Two}
\end{flalign}
The corresponding jump operators $\{\hat{J}_1, \hat{J}_2, \hat{J}_3\}$ are the same as in the single-peak case, see Eqs.~(\ref{eq:J1}-\ref{eq:J3}).
In the limit $\Delta \omega \rightarrow 0$ corresponding to single peak case, the rates~(\ref{eq:gamma1Two}-\ref{eq:gamma3Two}) reduce to those given by
 Eqs.~(\ref{eq:gamma1}-\ref{eq:gamma3}).

\begin{figure}[t]
\centering
\includegraphics[width=0.95 \columnwidth, height = 1.5 cm]{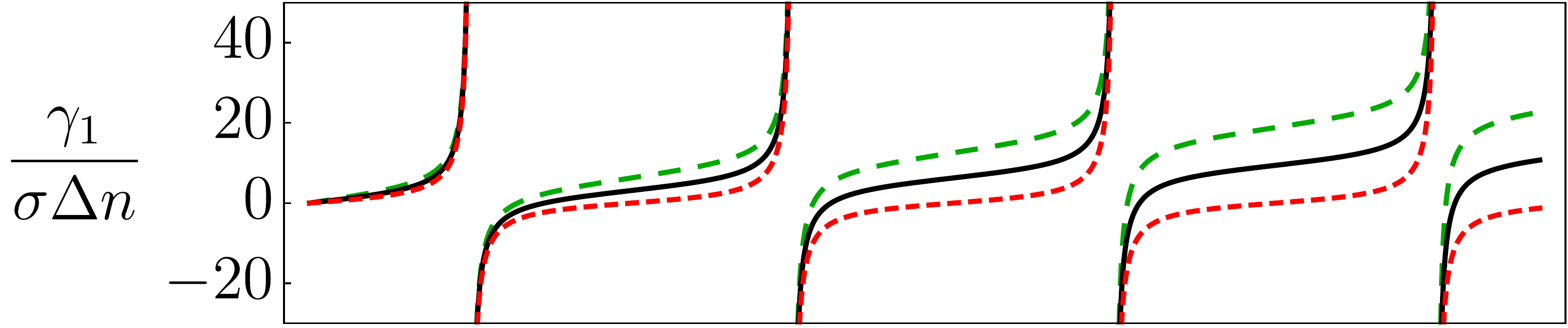}
\includegraphics[width=0.95 \columnwidth, height = 1.5 cm]{gamma2.pdf}
\includegraphics[width=0.95 \columnwidth ,height = 2.4 cm]{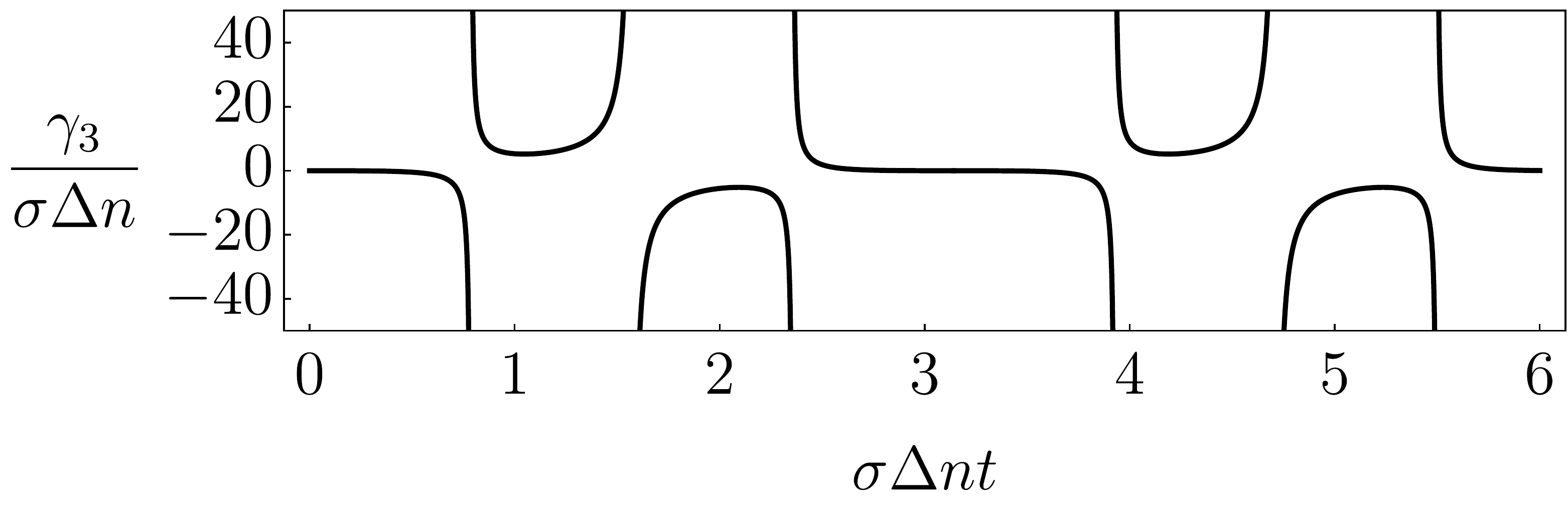}
\caption{(Color online) Decay rates as a function of normalized interaction time in the case of double-peak Gaussian frequency distribution. Dashed green when $K=-1$, solid black when $K=0$, and dot-dashed when $K=-1$. Here we set $\Delta \omega /\sigma=2$.}
\label{fig2}
\end{figure}

Figure~\ref{fig2} displays the rates for $K=-1,0,1$.
Dephasing rate $\gamma_2$ remains the same as in the single peak case. However, rate $\gamma_1$ -- corresponding to $\hat{J}_1$ including the sum of the local jump operators -- changes. The rate includes now an extra term, coming from the peak separation $\Delta \omega$,  
and an oscillatory part displaying negative values of the rate as a function of time. This also leads to non-Markovian dephasing dynamics which is not CP-divisible. It is even more striking that introducing the double-peak frequency structure, opens now an additional dephasing channel since the rate $\gamma_3$ is non-zero. Here, the corresponding jump operator $\hat{J}_3=\frac{1}{2}\hat{\sigma}_z \otimes \hat{\sigma}_z$ displays a joint bipartite structure, in contrast to local features of $\hat{J}_1$ and $\hat{J}_2$. This is an interesting observation since the system-environment interaction Hamiltonian is the same as before having only local interactions, see Eq.~(\ref{eq:HamiltonianTwoPhoton}), whilst the only change introduced was going from single- to double-peak structure of the initial bipartite environmental state. It is also worth noting that even though $\gamma_3$ is independent 
of $K$, its functional form is non-trivial since it contains the peak separation $\Delta \omega$ and trigonometric functions. 

There is a somewhat subtle mathematical point related to the behavior of rates $\gamma_1$ [Eq.~\eqref{eq:gamma1Two}] and $\gamma_3$ [Eq.~\eqref{eq:gamma3Two}] which needs an attention. Indeed, $\gamma_1(t)$ and $\gamma_3(t)$ diverge at isolated points of times. Subsequently, the corresponding dynamical maps are non-invertible at these points. According to the Eq.~\eqref{eq:MapBlochRep}, the generalized Bloch vector of the two-photon polarization state at time $t$ reads

\begin{equation}
\vec{r}(t)=\left(
\begin{array}{c}
 \frac{1}{2} \\
  \Gamma_0\cos(t\Delta\Omega/2)[\cos(t\Omega_0/2)r_2-\sin(t\Omega_0/2)r_3] \\
   \Gamma_0\cos(t\Delta\Omega/2)[\cos(t\Omega_0/2)r_3+\sin(t\Omega_0/2)r_2] \\
 r_4 \\
   \Gamma_0\cos(t\Delta\Omega/2)[\cos(t\Omega_0/2)r_5-\sin(t\Omega_0/2)r_6] \\
   \Gamma_0\cos(t\Delta\Omega/2)[\cos(t\Omega_0/2)r_6+\sin(t\Omega_0/2)r_5] \\
  \Gamma_-\cos(t\Delta \Omega)r_7 \\
  \Gamma_-\cos(t\Delta \Omega)r_8 \\
 r_9 \\
  \Gamma_+[\cos(t\Omega_0)r_{10}-\sin(t\Omega_0)r_{11}] \\
  \Gamma_+[\cos(t\Omega_0)r_{11}+\sin(t\Omega_0)r_{10}] \\
  \Gamma_0\cos(t\Delta\Omega/2)[\cos(t\Omega_0/2)r_{12}-\sin(t\Omega_0/2)r_{13}] \\
   \Gamma_0\cos(t\Delta\Omega/2)[\cos(t\Omega_0/2)r_{13}+\sin(t\Omega_0/2)r_{12}] \\
   \Gamma_0\cos(t\Delta\Omega/2)[\cos(t\Omega_0/2)r_{14}-\sin(t\Omega_0/2)r_{15}] \\
   \Gamma_0\cos(t\Delta\Omega/2)[\cos(t\Omega_0/2)r_{15}+\sin(t\Omega_0/2)r_{14}] \\
 r_{16} \\
\end{array}
\right),
\end{equation}
where we have defined $\Gamma_0=\exp[-\sigma^2 \Delta n^2 t^2/2]$, $\Gamma_{\pm}=\exp[-\sigma^2 \Delta n^2 t^2(1\pm K)]$, $ \Delta\Omega= \Delta n \Delta \omega$, $\Omega_0=\Delta n \omega_0$, and $\vec{r}(0)=(1/2,r_2,r_3,...,r_{16})^T$ is the initial Bloch vector. One can check that all of the different initial vectors (states) that share the same values of $r_4,r_7,r_8,r_9,r_{10},r_{11}$, and $r_{16}$ are mapped to the same vector (state) at $t=\pi /\Delta \Omega$. This many to one nature of the map --  at these isolated times -- makes it non-invertible. Although all the trajectories corresponding to the aforementioned initial vectors end up together at the isolated points, it is evident that they continue their different paths immediately after this. This can be seen in the following way. Consider the generator of the master equation in matrix form and its action on the generalized Bloch vector.
We see that while some rates diverge at certain points in time, it is precisely at these points that the generalized Bloch vector components -- with which the rates get multiplied -- all go to zero. In more detail, we have
\begin{equation}
\frac{d}{dt} r_{\alpha}(t)=\sum_{\beta}[\hat{\mathcal{L}}_t]_{\alpha \beta} \, r_{\beta}(t),
\end{equation}
therefore, the product of the divergent rate with the zero value component leads to a finite rate of change of the Bloch vector which allows us to continue propagation of each state forward in time. Accordingly, following the trajectories immediately before they unite at a single point, lets us identify each one of them immediately after that, when they separate again. We see therefore that in spite of the divergences in the rates, the master equation we have obtained describes the dephasing evolution of the two-photon polarization state in meaningful way.
It is also worth noting that the divergent decoherence rates in master equations have appeared in earlier literature many times, e.g., in the prominent resonant Jaynes-Cummings model~\cite{BreuerBook_2002}.
 
%%%%%%%%%%%%%%%%%%%%%%%%%%%%% 
%%%%%%%%%%%%%%%%%%%%%%%%%%%%%
\section{Single-photon dephasing with initial polarization-frequency correlations} \label{sec4}

We described above how initial correlations between the composite environmental states influence the generator of the dynamical map and the corresponding master equation for photonic dephasing. In this section we continue with initial correlations but 
take a different perspective by considering non-factorized initial system-evironmental state for single qubit.
This is motivated by the recent observation that initial system-environment correlations can be exploited for arbitrary control of single qubit dephasing~\cite{Lyyra_2018}. We revisit this problem and obtain new insight by exploiting the very recently developed general method of \textit{bath positive decomposition} ($\mathrm{B+}$ decomposition) ~\cite{Wiseman_2019}. 
In general,  presence of initial system-environment correlations implies that the open system evolution is not described by a CP dynamical map   \cite{pechukas94,Alicki_1995,pechukas95,jordan04,shaji05a,Linta-Shaji}. However, $\mathrm{B+}$ decomposition method allows to treat this case with a set of CP maps, where each term of the decomposition is evolved over time with its individual CP map~\cite{Wiseman_2019}. 

\subsection{Preliminaries on $\mathrm{B+}$ decomposition for initially correlated system-environment state}
Following~\cite{Wiseman_2019} we begin by considering arbitrary system-environment state -- in the corresponding Hilbert space $\mathcal{H}=\mathcal{H}_S\otimes \mathcal{H}_E$ -- and write it as 
\begin{equation}\label{eq:BplusDecom}
\hat{\rho}_{SE}(0)=\sum_{\alpha}w_{\alpha}\hat{Q}_{\alpha}\otimes \hat{\rho}_{\alpha}.
\end{equation} 
Here, $\{\hat{Q}_{\alpha}\}$ forms a basis (possibly overcomplete) for operators on $\mathcal{H}_S$ and $\{\hat{\rho}_{\alpha}\}$ are valid environmental density operators on $\mathcal{H}_E$. Note that $\hat{Q}_{\alpha}$ need not be positive or trace orthogonal, so they may not constitute proper density matrices on the system Hilbert space. However when the initial state is factorized, this summation reduces to a single term $\hat{\rho}_{SE}(0)=\hat{\rho}_S(0)\otimes \hat{\rho}_E(0)$ corresponding to reduced states of the open system and environment, respectively. In general, number of terms in this summation is restricted by $1\leq N\leq d^2$ where $d$ is the dimension of the system Hilbert space \cite{Wiseman_2019}. All the information about initial state of the open system is incorporated in the weights $w_{\alpha}$, such that $\hat{\rho}_S(0)=\mathrm{Tr}_E[\hat{\rho}_{SE}(0)]=\sum w_{\alpha}\hat{Q}_{\alpha}$. Although $\hat{Q}_{\alpha}$ may not be legitimate density operators for the open system, those expressed by $\hat{\rho}_{\alpha}$ are valid density operators for the environment. This means that the factorized form of the terms in \eqref{eq:BplusDecom} allows to write the dynamics of the open system state as the weighted sum of legitimate CP-maps acting on $\hat{Q}_{\alpha}$. In more detail, if the total system-environment evolves due to a unitary operator $\hat{U}(t)$, one has
\begin{eqnarray}
\hat{\rho}_S(t)&&=\sum_{\alpha}w_{\alpha}\mathrm{Tr}_E[\hat{U}(t)(\hat{Q}_{\alpha} \otimes \hat{\rho}_{\alpha})\hat{U}(t)^{\dagger}] \nonumber
\\
&&=\sum_{\alpha}w_{\alpha}\hat{\Phi}^{(\alpha)}_t[\hat{Q}_{\alpha}],
\end{eqnarray}
where
\begin{equation}\label{eq:BPlusDecMap}
\hat{\Phi}_t^{(\alpha)}[\cdot]:=\mathrm{Tr}_E[\hat{U}(t)(\cdot\otimes\hat{\rho}_{\alpha})\hat{U}(t)^{\dagger}].
\end{equation}
Since all maps of the form given in Eq.\eqref{eq:BPlusDecMap} are CP, all previous tools for studying CP-maps are applicable here. In particular, one can investigate properties of each CP-map $\Phi^{(\alpha)}_t$ and see how they are connected to the presence of initial correlations.

For example, consider single qubit dynamics in the presence of initial system-environment correlations~\cite{Wiseman_2019} . Using completeness of Pauli sigma basis $\{I_2,\hat{\sigma}_x,\hat{\sigma}_y,\hat{\sigma}_z\}$, we have 
\begin{equation}\label{eq:BplusDecQubit}
\hat{\rho}_{SE}(0)=\sum_{\alpha=0,x,y,z} w_{\alpha} \hat{Q}_{\alpha}\otimes \hat{\rho}_{\alpha},
\end{equation}
in which
\begin{eqnarray}
&&\hat{Q}_0=\frac{1}{2}(\hat{I}_2 - \hat{\sigma}_x - \hat{\sigma}_y - \hat{\sigma}_z),\label{eq:Q0}\\
&&\hat{Q}_{\alpha}=\frac{1}{2}\hat{\sigma}_{\alpha} \quad for \; \alpha = x,y,z,\label{eq:Qalpha}
\end{eqnarray}
and
\begin{eqnarray}
&&\hat{\rho}_0=\mathrm{Tr}_S[\hat{\rho}_{SE}(0)]=\hat{\rho}_E(0),\label{rho1}
\\
&&\hat{\rho}_{\alpha}=\frac{\mathrm{Tr}_S[((\hat{I}_2+\hat{\sigma}_{\alpha})\otimes \hat{I}_E)\hat{\rho}_{SE}(0)]}{w_{\alpha}},\label{rho2}
\end{eqnarray}
with $w_0=1$, and $w_{\alpha}=\mathrm{Tr}[((\hat{I}_2+\hat{\sigma}_{\alpha})\otimes \hat{I}_E)\hat{\rho}_{SE}(0)]$ for ${\alpha}=x,y,z$.
We exploit these generic expressions below.

\subsection{Initial polarization-frequency correlation and $\mathrm{B+}$ decomposition for single photon}

We consider initial polarization-frequency correlations by following the recent results and experimental work on generating, in principle, arbitrary single-photon dephasing dynamics~\cite{Lyyra_2018}. Generic initial polarization-frequency state can be written as
\begin{eqnarray}\label{eq:InStateCorrelated}
\ket{\psi(0)}_{SE}=&& C_v\ket{v}\otimes\int d \omega g(\omega)\ket{\omega}\nonumber
\\
&&\quad+C_h\ket{h}\otimes\int d \omega g(\omega)\mathrm{e}^{i \theta(\omega)}\ket{\omega},
\end{eqnarray}
where $\vert C_h \vert ^2+\vert C_v \vert ^2=1$ and $\int d \omega \vert g(\omega ) \vert ^2=1$.
Above, the crucial ingredient is the frequency dependent initial phase $\theta(\omega)$ for the component including the polarization $h$.
If $\theta(\omega)$ is a constant function, then there are no initial system-environment correlations. However, controlling the non-constant functional form of $\theta(\omega)$
allows to control the initial correlations and their amount.
  
When the initial state evolves according to the interaction Hamiltonian in Eg.~\eqref{eq:Hamiltonian}, the reduced polarization state at time $t$ is
\begin{equation}\label{eq:MapInCorrel}
\hat{\rho}(t)=\begin{pmatrix}
\vert C_h \vert ^2 & \kappa(t)C_h C_v^{*} \\
\kappa(t)^{*}C_h^{*} C_v & \vert C_v \vert ^2
\end{pmatrix},
\end{equation}
where the decoherence function is 
\begin{equation}\label{eq:decFuncCorrel}
\kappa(t)=\int d \omega \vert g(\omega) \vert ^2  \mathrm{e}^{i \theta(\omega)}\mathrm{e}^{-i \Delta n \omega t}.
\end{equation}
Note that in addition to the frequency probability distribution $|g(\omega)|^2$, one can now use also $\theta(\omega)$, and subsequent initial correlations, to control the dephasing dynamics.

The dynamics given by Eqs.~(\ref{eq:MapInCorrel}-\ref{eq:decFuncCorrel}) can be equivalently formulated by using the $\mathrm{B}+$ decomposition. 
Considering the initial total state in Eq.~\eqref{eq:InStateCorrelated}, and applying the $\mathrm{B}+$ decomposition along Eq.~\eqref{eq:BplusDecQubit} and  Eq.~\eqref{rho1}-\eqref{rho2}, we obtain environmental terms
\begin{widetext}
\begin{eqnarray}
&&\hat{\rho}_0=\int d\omega \int d\omega'\; g(\omega)g(\omega')^{*}\;\big(\vert C_h\vert ^2 \mathrm{e}^{i[\theta(\omega)-\theta(\omega')]}+\vert C_v\vert ^2\big)\ket{\omega}\bra{\omega'},
\\
&&\hat{\rho}_x=\frac{1}{w_x}\int d\omega \int d\omega'\; g(\omega)g(\omega')^{*}\;\big(\vert C_h\vert ^2 \mathrm{e}^{i[\theta(\omega)-\theta(\omega')]}+\vert C_v\vert ^2+C_h C_v^{*}\mathrm{e}^{i\theta(\omega')}+C_v C_h^{*}\mathrm{e}^{-i\theta(\omega)}\big)\ket{\omega}\bra{\omega'},
\\
&&\hat{\rho}_y=\frac{1}{w_y}\int d\omega \int d\omega' \;g(\omega)g(\omega')^{*}\;\big(\vert C_h\vert ^2 \mathrm{e}^{i[\theta(\omega)-\theta(\omega')]}+\vert C_v\vert ^2+i C_h C_v^{*}\mathrm{e}^{i\theta(\omega')}- i C_v C_h^{*}\mathrm{e}^{-i\theta(\omega)}\big)\ket{\omega}\bra{\omega'},
\\
&&\hat{\rho}_z=\int d\omega \int d\omega'\; g(\omega)g(\omega')^{*}\;\ket{\omega}\bra{\omega'},
\end{eqnarray}
\end{widetext}
with weights
\begin{eqnarray}
&&w_x=1+2\int d\omega \vert g(\omega)\vert ^2 \Re{[C_v C_h^{*}\mathrm{e}^{-i\theta(\omega)}]},
\\
&&w_y=1+2\int d\omega \vert g(\omega)\vert ^2 \Im{[C_v C_h^{*}\mathrm{e}^{-i\theta(\omega)}]},
\\
&&w_z=2\vert C_h\vert ^2.
\end{eqnarray}
Each specific term of the $\mathrm{B}+$ decomposition is related to a frequency state ($\hat{\rho}_{\alpha}$) above, and acts on its own input system operator $\hat{Q}_{\alpha}$, see Eqs~(\ref{eq:Q0}-\ref{eq:Qalpha}). In the current case, we can combine the contributions of $\hat{\rho}_0$ and $\hat{\rho}_z$ to simplify the decomposition into only three terms. Subsequently, polarization density matrix at time $t$ is given by  
\begin{eqnarray}\label{eq:MapInCorrThreeTerms}
\hat{\rho}(t)&=&\frac{1}{2}\begin{pmatrix} \nonumber
w_z & \kappa_0(t)(i-1) \\
\kappa_0(t)^{*}(-i-1) & 2-w_z
\end{pmatrix}
\\
&&+\frac{1}{2} w_x\begin{pmatrix} \nonumber
0 & \kappa_x(t) \\
\kappa_x(t)^{*} & 0
\end{pmatrix}
\\
&&+\frac{1}{2} w_y\begin{pmatrix}
0 & -i \kappa_y(t) \\
i \kappa_y(t)^{*} & 0
\end{pmatrix},
\end{eqnarray}
where the three different decoherence functions are given by
\begin{flalign}
&\kappa_0(t)=\int d \omega \vert g(\omega)\vert ^2 \mathrm{e}^{-i \Delta n \omega t}, \label{d1}
\\
&\kappa_x(t)=\frac{\int d \omega \vert g(\omega)\vert ^2 (1+2 \Re{[C_v C_h^{*}\mathrm{e}^{-i\theta(\omega)}]})\mathrm{e}^{-i \Delta n \omega t}}{w_x}, \label{d2}
\\
&\kappa_y(t)=\frac{\int d \omega \vert g(\omega)\vert ^2 (1+2 \Im{[C_v C_h^{*}\mathrm{e}^{-i\theta(\omega)}]})\mathrm{e}^{-i \Delta n \omega t}}{w_y}.\label{d3}
\end{flalign}
It is interesting to note here that the decoherence function $\kappa_0$ is independent of $\theta(\omega)$ and actually corresponds directly to the case when there are no initial polarization-frequency correlations.
The other two functions, $\kappa_x$ and $\kappa_y$, depend also on $\theta(\omega)$ and describe in detail how the initial correlations change the dephasing dynamics.

%%%%%%%%%%%%%%%%%%%%%%%%%%%%%
\begin{figure}[t]
\centering
\includegraphics[width=0.8 \columnwidth, height = 1.5 cm]{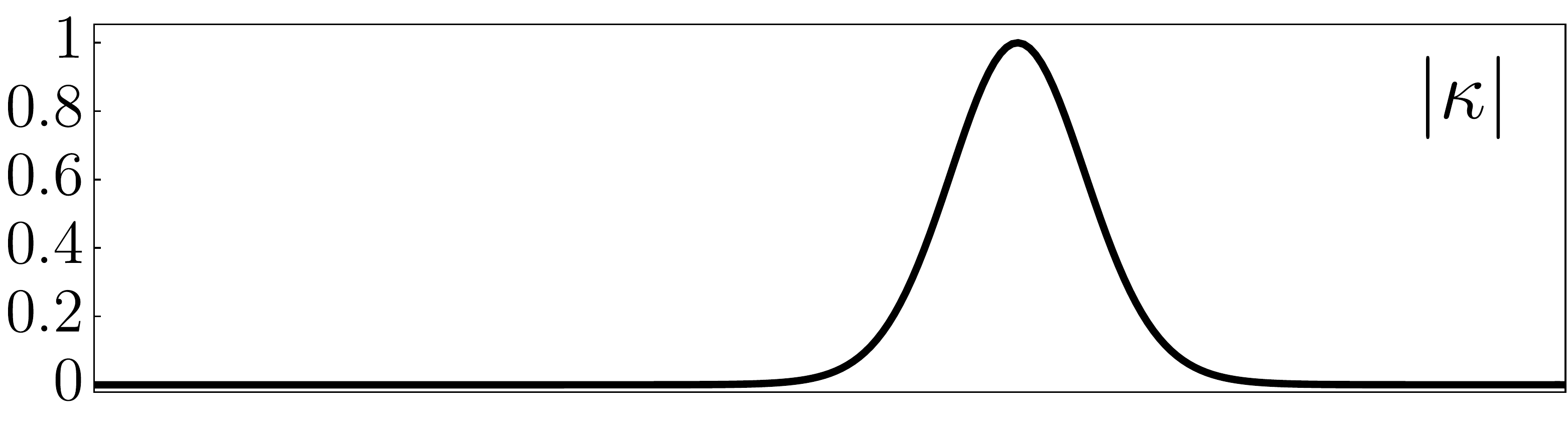}
\includegraphics[width=0.8 \columnwidth, height = 1.5 cm]{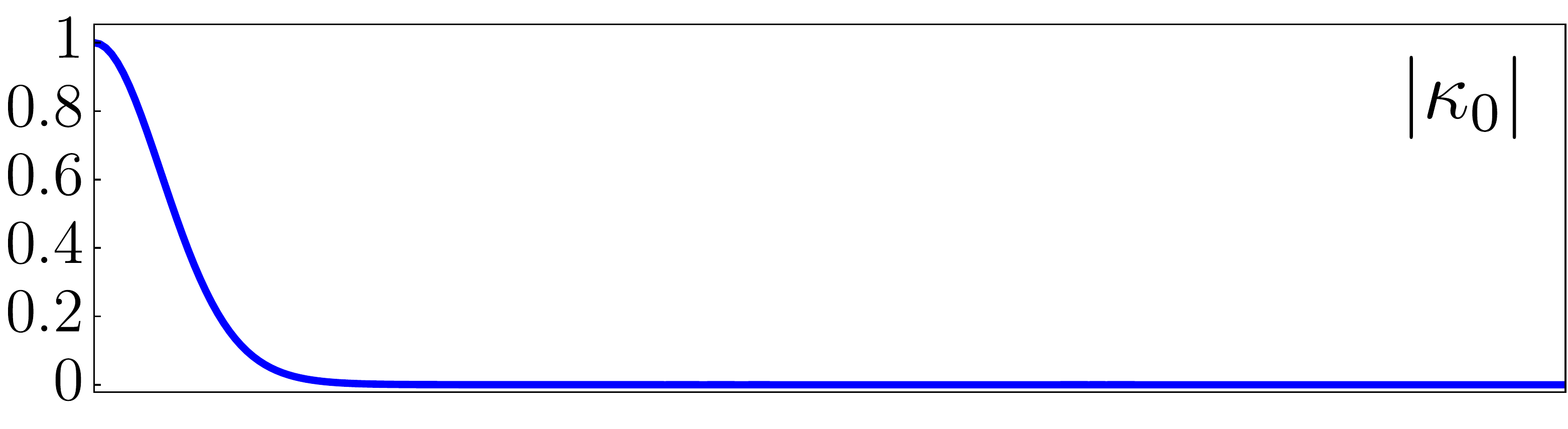}
\includegraphics[width=0.8 \columnwidth, height = 1.5 cm]{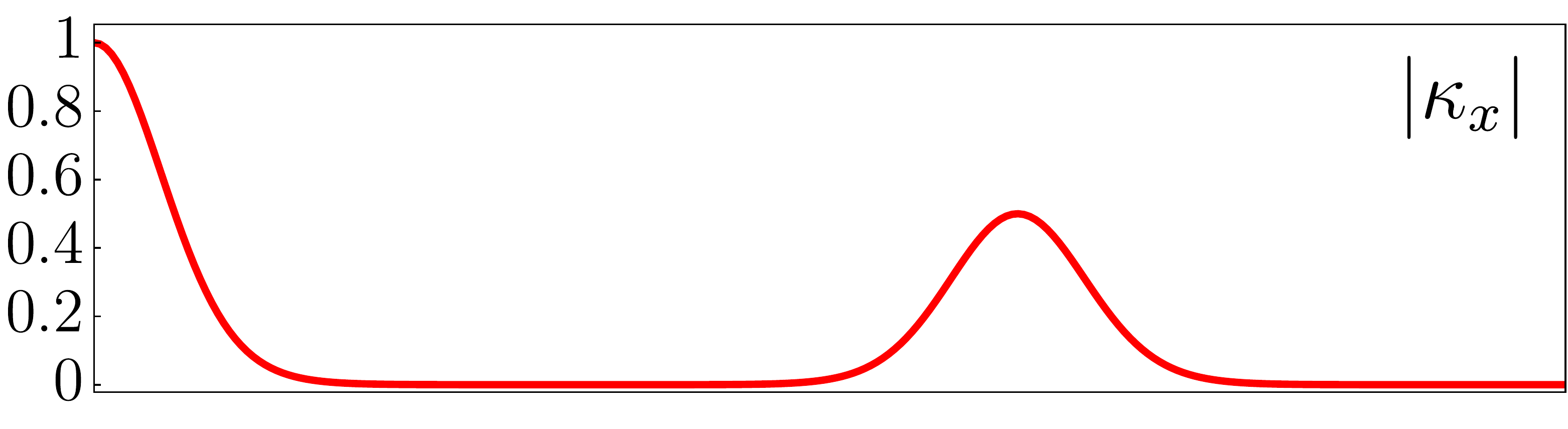}
\includegraphics[width=0.8 \columnwidth ,height = 2 cm]{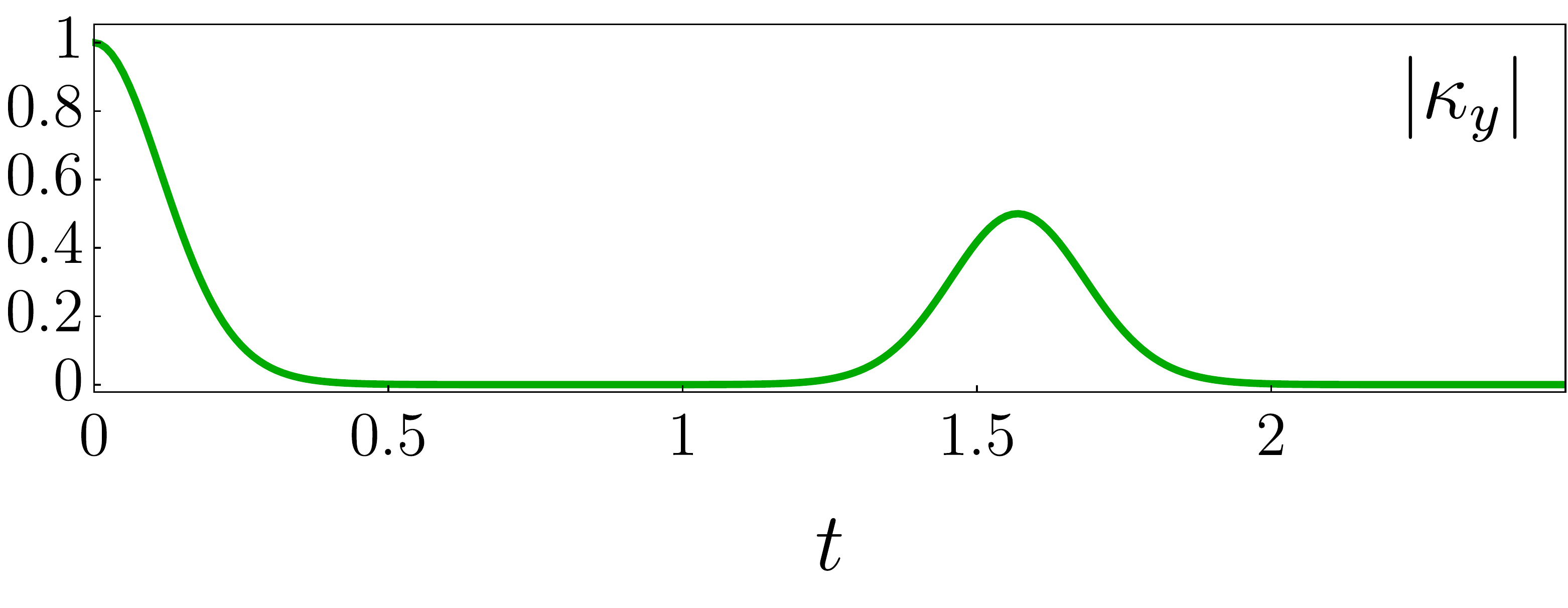}
\caption{(Color online) Non-positive map decoherence functions. Magnitudes of the original decoherence function ($\kappa$) and $\mathrm{B}+$ decomposition decoherence functions ($\kappa_0, \kappa_x, \kappa_y$) as a function of time. We set $C_h=C_v=1/\sqrt{2}$.}\label{fig3}
\end{figure}
%%%%%%%%%%%%%%%%%%%%%%%%%%%%%

It is also interesting to compare Eq.~\eqref{eq:MapInCorrThreeTerms} with the $\mathrm{B+}$ decomposition for generic dephasing dynamics of a qubit coupled to a Bosonic bath, when qubit and bath are initially correlated \citep{Wiseman_2019}. The total Hamiltonian of the qubit and the Bosonic bath reads
\begin{equation}
\hat{H}=\omega_q \; \hat{\sigma}_z+\sum_i \; \omega_i \hat{b}^{\dagger}_i \hat{b}_i+\hat{\sigma}_z\otimes \sum_i g_i (\hat{b}^{\dagger}_i+\hat{b}_i),
\end{equation}
where $\omega_q$ is the qubit's energy level separation (in $\ket{0}$, $\ket{1}$ basis), $\hat{b}^{\dagger}_i$ and $\hat{b}_i$ are bath mode creation and annihilation operators respectively, and $g_i$ is the coupling strength. Employing the $\mathrm{B+}$ decomposition, dynamics of the off-diagonal element of the qubit's density matrix in the interaction picture reads \citep{Wiseman_2019}
\begin{equation}\label{eq:dephasScale}
\bra{0}\rho_S(t)\ket{1}=\sum_{\alpha}\;w_{\alpha} \bra{0}\hat{Q}_{\alpha}\ket{1}\chi_{\hat{\rho}_{\alpha}}(\vec{\xi}_t),
\end{equation}
where $\chi_{\hat{\rho}_{\alpha}}(\vec{\xi}_t)=\mathrm{Tr}_B[\hat{\rho}_{\alpha}\hat{D}(\vec{\xi}_t)]$ is the Wigner characteristic function of the bath state $\hat{\rho}_{\alpha}$. Above $\vec{\xi}_t=(\xi_1(t),\xi_2(t),...)$ with
\[\xi_j(t)=2g_j \bigg(\frac{1-\e^{i\omega_j t}}{\omega_j}\bigg) ,\]
and $\hat{D}(\vec{\xi}_t)=\exp(\sum_i \xi_i \hat{b}^{\dagger}_i+\xi^{*}\hat{b}_i)$ is the Glauber displacement operator. The comparison between 
Eqs.~\eqref{eq:MapInCorrThreeTerms} and \eqref{eq:dephasScale} reveals that the decoherence functions in our photonic model -- corresponding to integral transformations of the frequency probability distribution and frequency dependent phase $\theta(\omega)$ --   play the exact same role as the characteristic functions in the dephasing dynamics of a qubit coupled to Bosonic bath.

Let us go back to the photonic model and see in detail, for some examples, what is the relation between the orginal decoherence function \eqref{eq:decFuncCorrel} and those appearing in the $\mathrm{B}+$ decomposition in Eqs.~(\ref{d1}-\ref{d3}). In particular, we consider the similar cases as used in~\citep{Lyyra_2018} to demonstrate arbitrary control of dephasing dynamics. These include a non-positive map, Markovian, non-Markovian, and coherence trapping dynamics.
In all of the cases below, the frequency distributions used and the values for $\theta(\omega)$ are similar to those considered in Ref.~\citep{Lyyra_2018}, respectively.

%%%%%%%%%%%%%%%%%%%%%%%%%%%%%
\begin{figure}[t]
\centering
\includegraphics[width=0.83 \columnwidth, height = 1.5 cm]{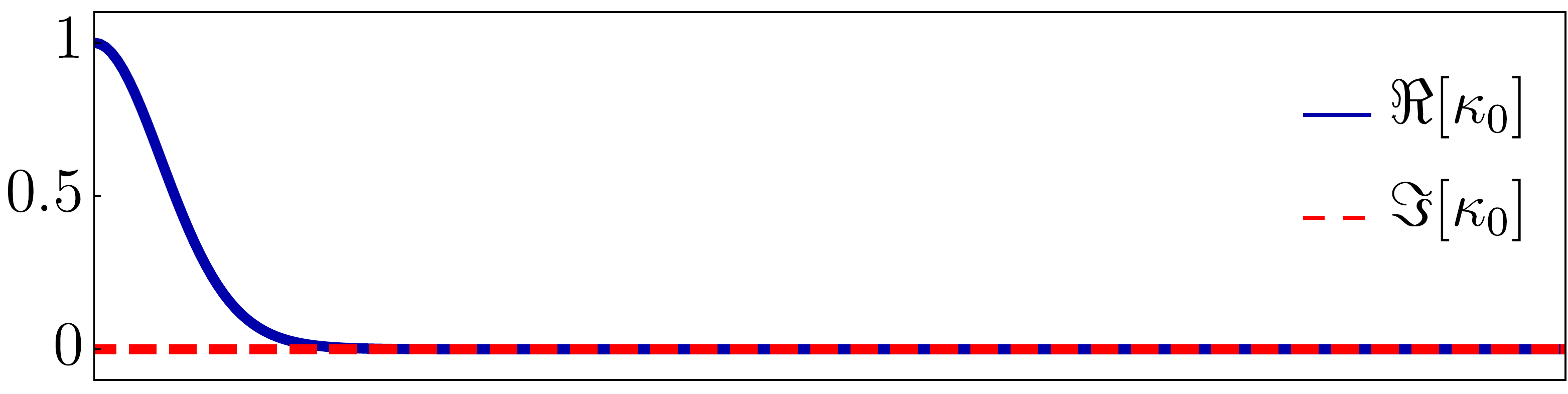}
\includegraphics[width=0.83 \columnwidth, height = 1.5 cm]{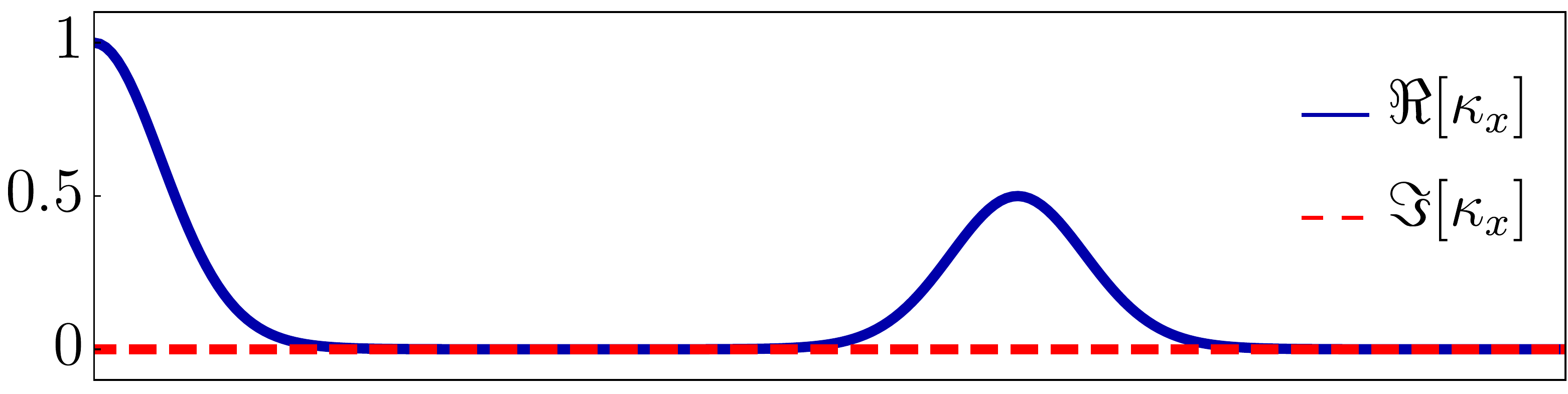}
\includegraphics[width=0.83 \columnwidth, height = 2 cm]{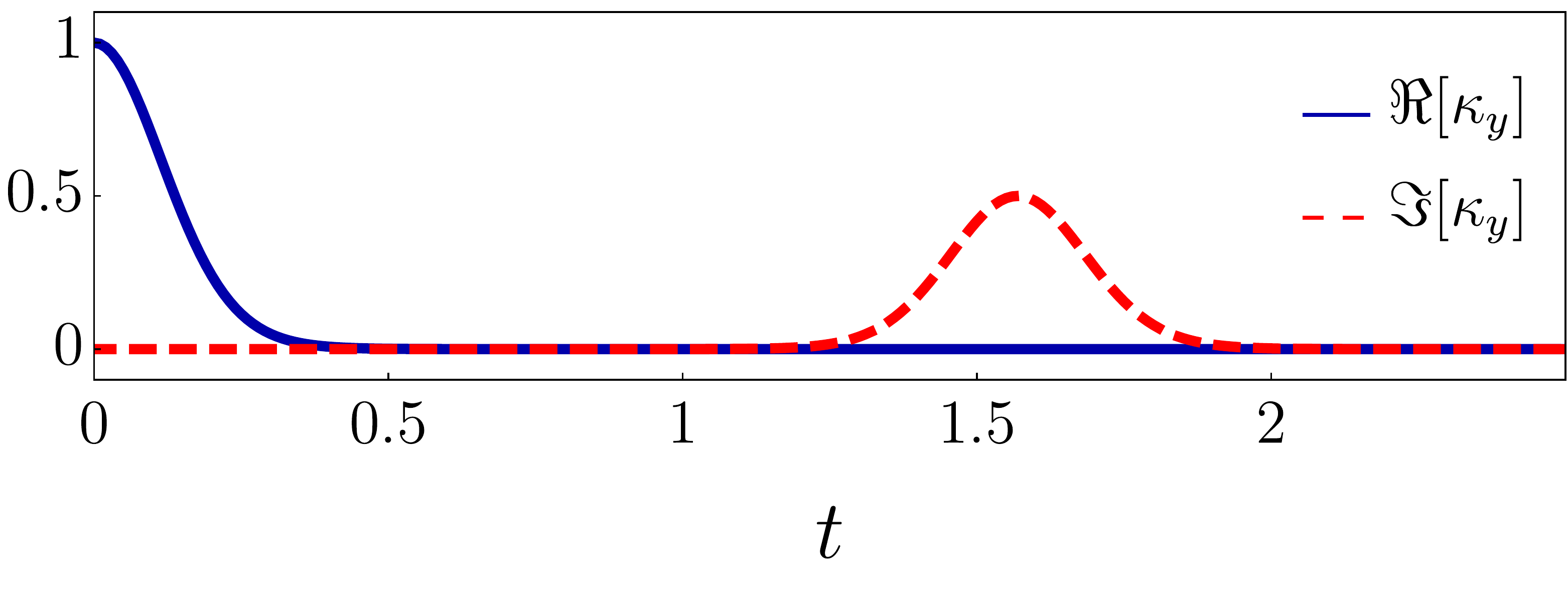}
\caption{(Color online) Non-positive map decoherence functions. Real and imaginary parts of the original decoherence function ($\kappa$) and $\mathrm{B}+$ decomposition decoherence functions ($\kappa_0, \kappa_x, \kappa_y$) as a function of time. We set $C_h=C_v=1/\sqrt{2}$.}
\label{fig4}
\end{figure} 
%%%%%%%%%%%%%%%%%%%%%%%%%%%%%

Figure \ref{fig3} shows the magnitude of various decoherence functions for the case of a non-positive (NP) map, i.e., $\kappa (t) > \kappa(0)$. 
 It is easy to check that the off-diagonal term of the density matrix is obtained from $\rho_{hv}(t)=(\kappa_0(t)(i-1)+w_x \kappa_x(t)-iw_y \kappa_y(t))/2$, and equivalently from $\rho_{hv}(t)=\kappa(t)C_h C_v^{*}$. Thereby, it is evident  that if $\kappa_0(t)=\kappa_x(t)=\kappa_y(t)=0$, for some $t>0$, then $\kappa(t)=0$. However, the reverse statement does not always hold. Instead, one can show that whenever $w_x=w_y=1$, then having identical decoherence functions, $\kappa_0(t)=\kappa_x(t)=\kappa_y(t)$, is sufficient to have zero coherence, i.e., $\kappa(t)=0$. This is an interesting result making a link between properties of the CP-maps obtained in $\mathrm{B+}$ decomposition and  the original non-positive map. In fact, the case discussed in Fig. \ref{fig3} demonstrates this situation.
This is even more evident when considering the real and imaginary parts of the decoherence functions explicitly, see Fig.~\ref{fig4}. 
One can see that the three decoherence functions $\kappa_0$, $\kappa_x$, and $\kappa_y$ are identical when the interaction time is short. Therefore, since we also have $w_x = w_y = 1$, the decoherence function $\kappa(t)$ has zero value in this regime.

\begin{figure}[t]
\centering
\includegraphics[width=0.8 \columnwidth, height = 1.5 cm]{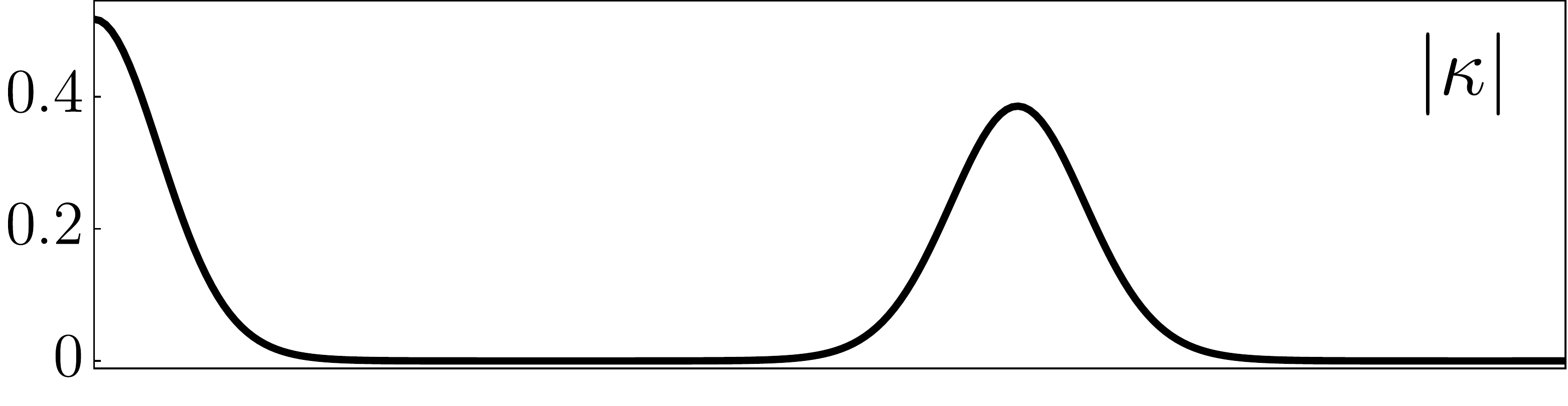}
\includegraphics[width=0.8 \columnwidth, height = 1.5 cm]{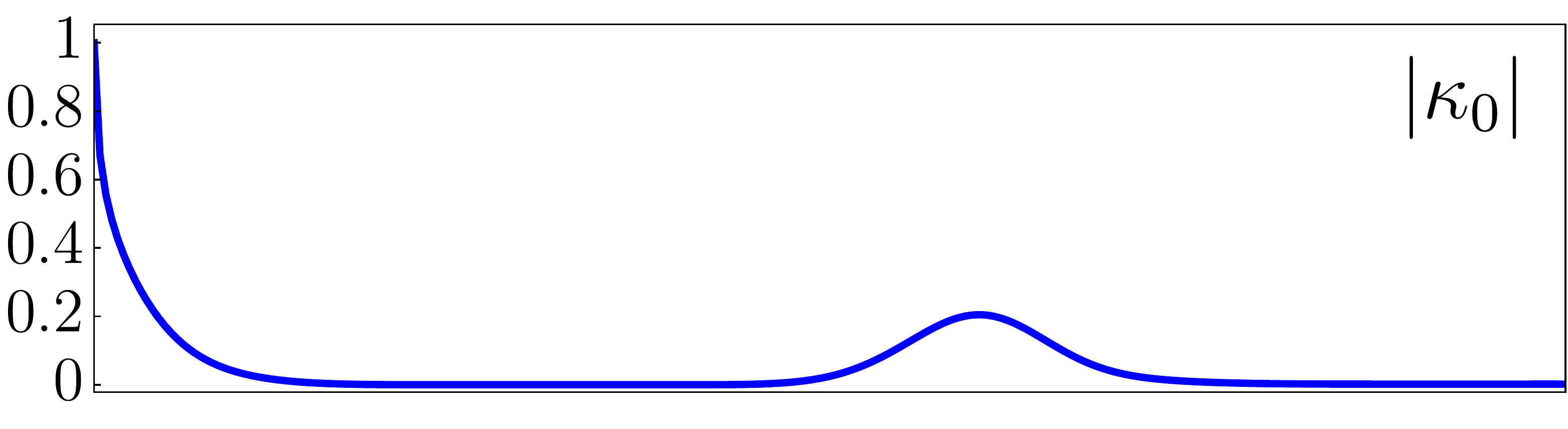}
\includegraphics[width=0.8 \columnwidth, height = 1.5 cm]{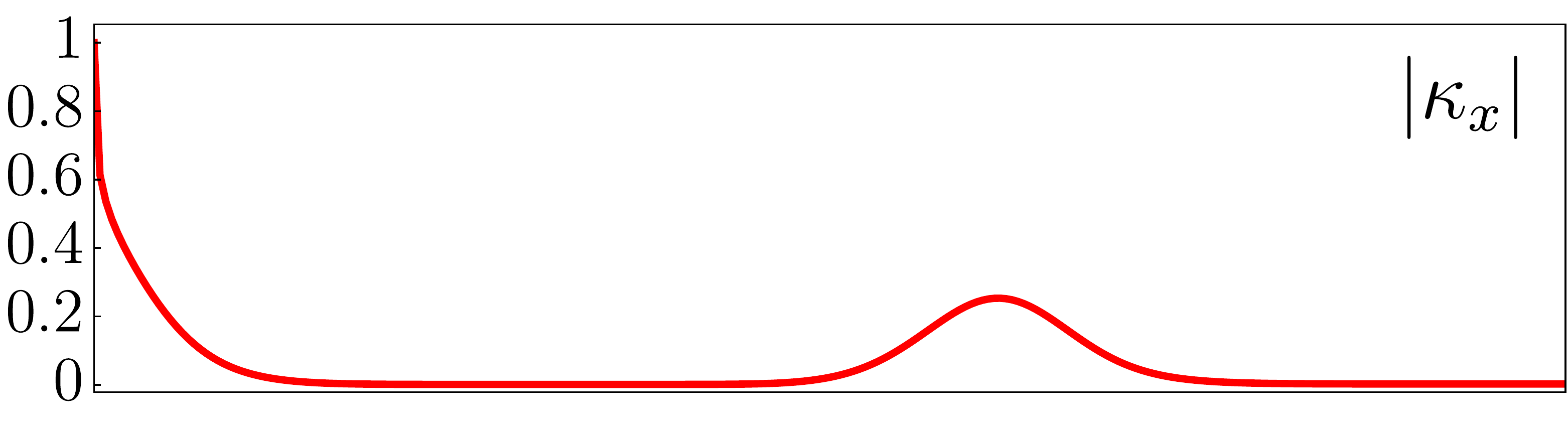}
\includegraphics[width=0.8 \columnwidth ,height = 2 cm]{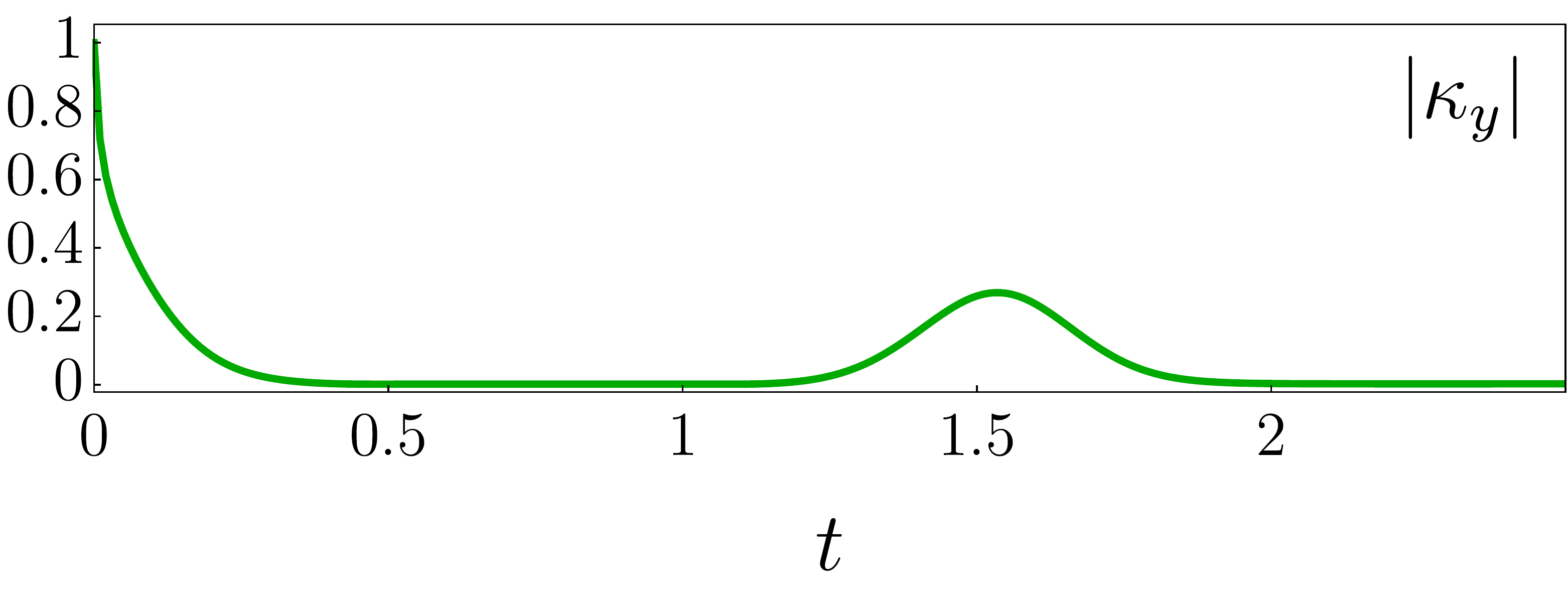}
\caption{(Color online) Non-Markovian dynamics. Magnitudes of the original decoherence function ($\kappa$) and $\mathrm{B}+$ decomposition decoherence functions ($\kappa_0, \kappa_x, \kappa_y$) as a function of time. We set $C_h=C_v=1/\sqrt{2}$.}\label{fig5}
\end{figure}

\begin{figure}[t]
\centering
\includegraphics[width=0.8 \columnwidth, height = 1.5 cm]{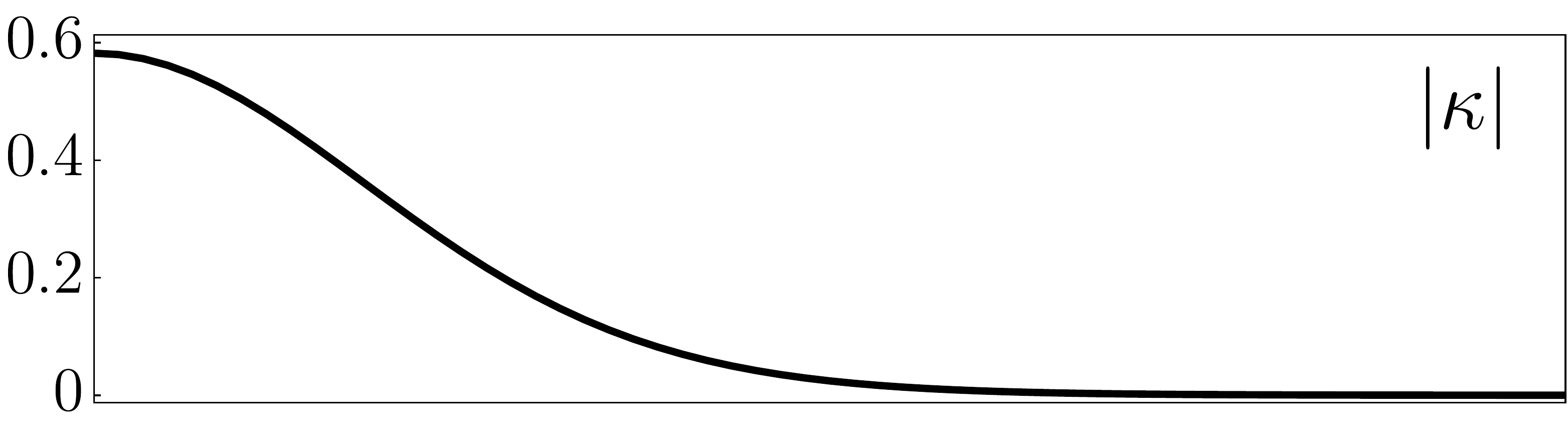}
\includegraphics[width=0.8 \columnwidth, height = 1.5 cm]{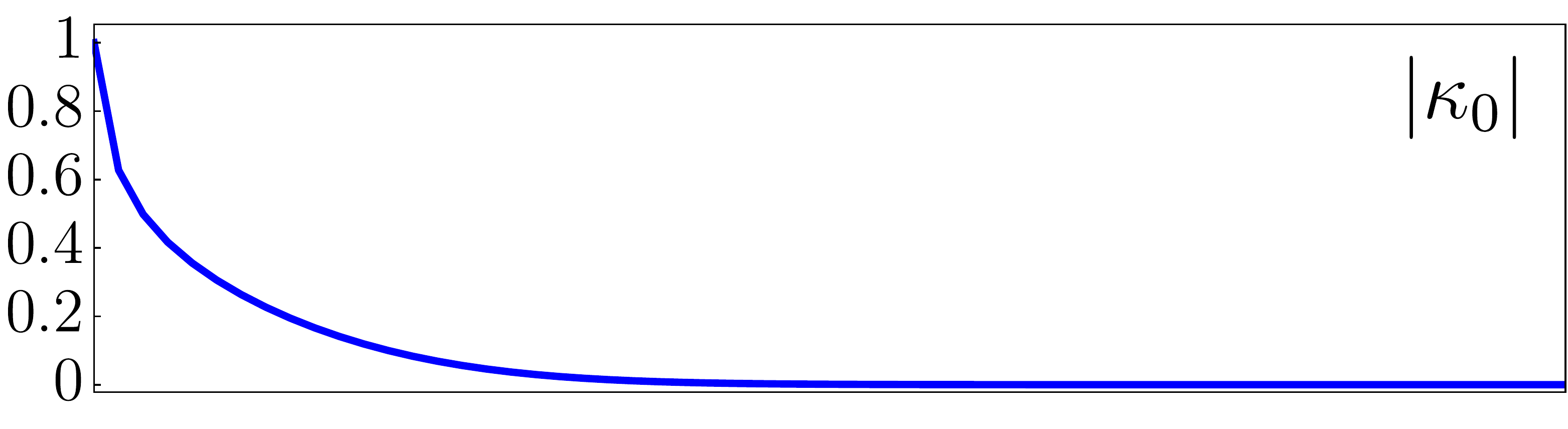}
\includegraphics[width=0.8 \columnwidth, height = 1.5 cm]{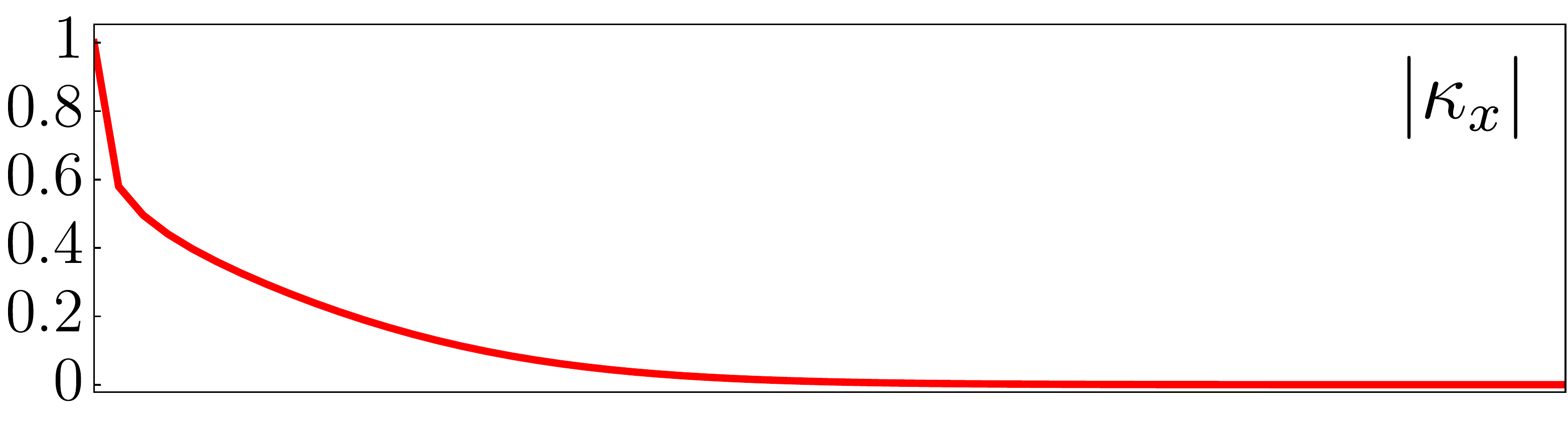}
\includegraphics[width=0.8 \columnwidth ,height = 2 cm]{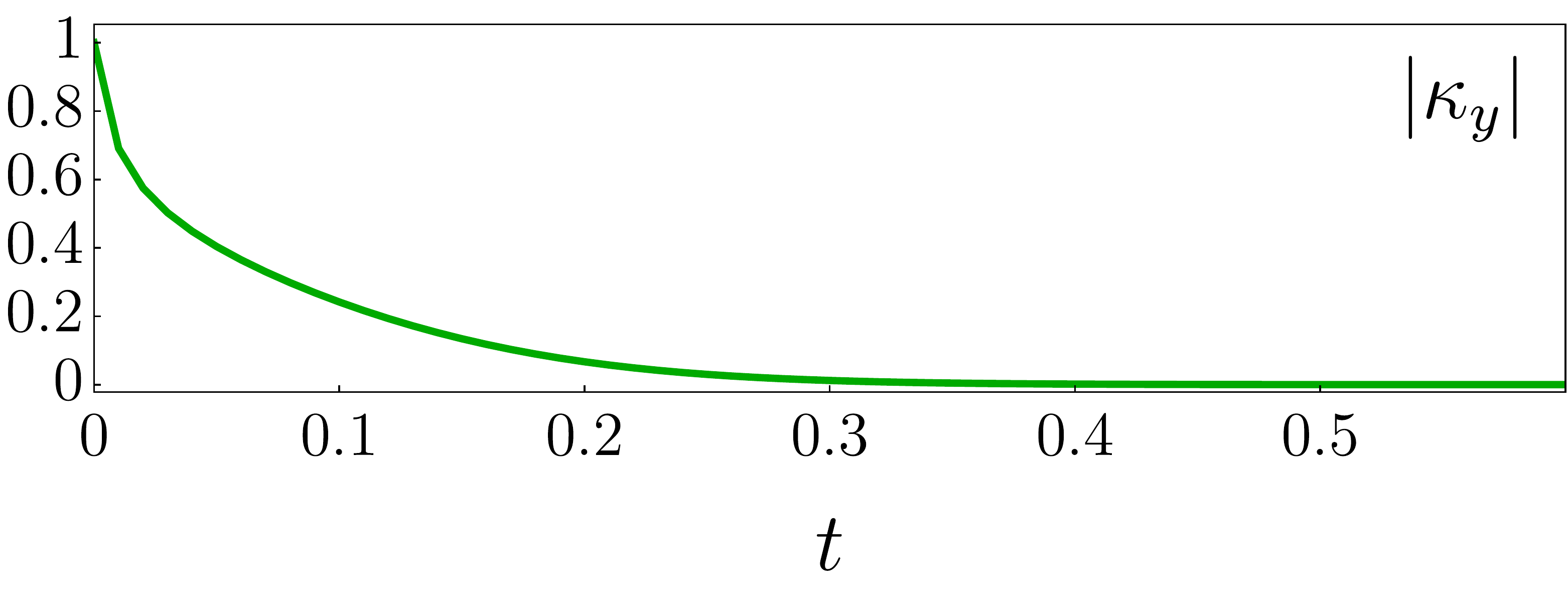}
\caption{(Color online) Markovian dynamics. Magnitudes of the original decoherence function ($\kappa$) and $\mathrm{B}+$ decomposition decoherence functions ($\kappa_0, \kappa_x, \kappa_y$) as a function of time. We set $C_h=C_v=1/\sqrt{2}$.}\label{fig6}
\end{figure}

\begin{figure}[t]
\centering
\includegraphics[width=0.8 \columnwidth, height = 1.5 cm]{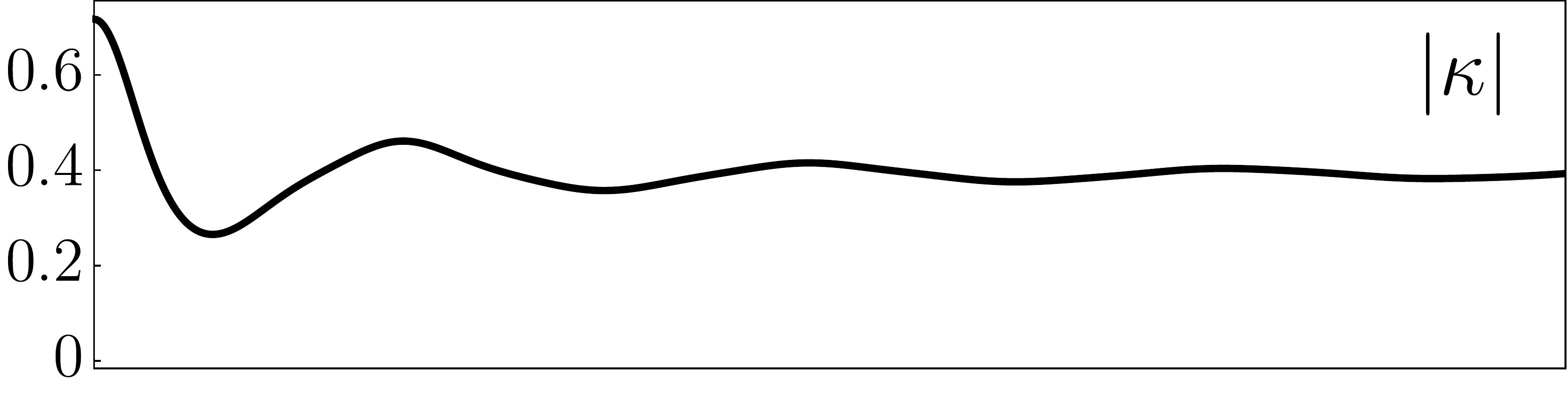}
\includegraphics[width=0.8 \columnwidth, height = 1.5 cm]{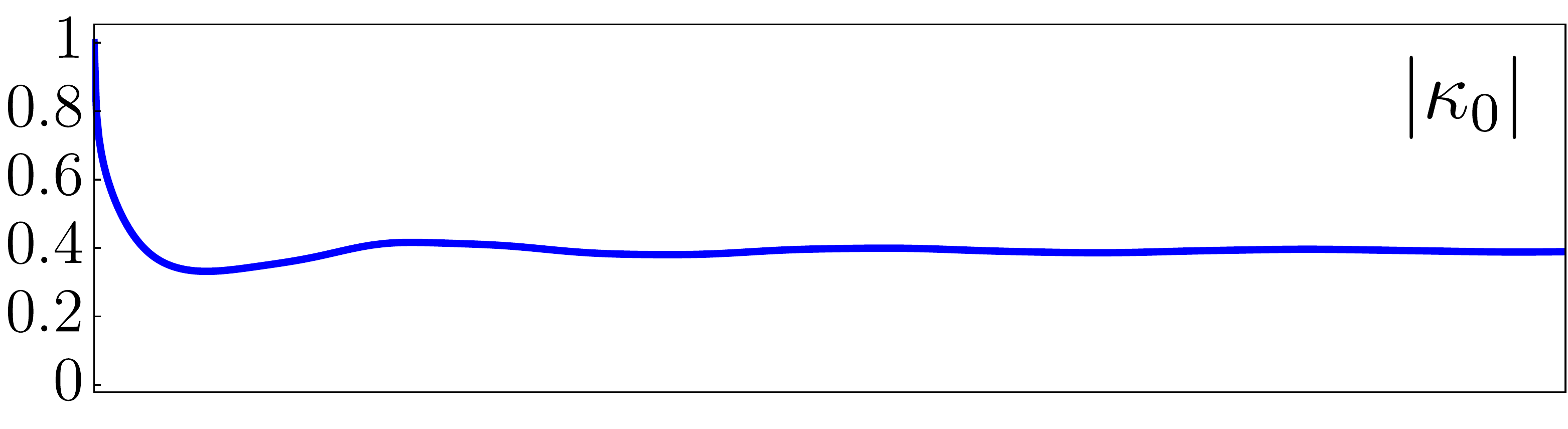}
\includegraphics[width=0.8 \columnwidth, height = 1.5 cm]{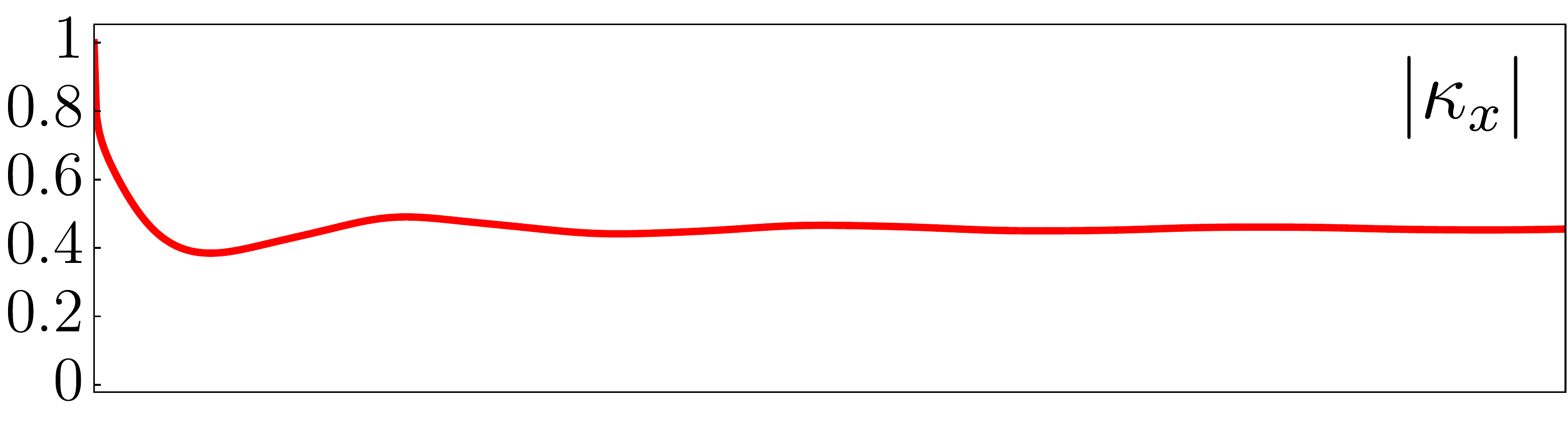}
\includegraphics[width=0.8 \columnwidth ,height = 2 cm]{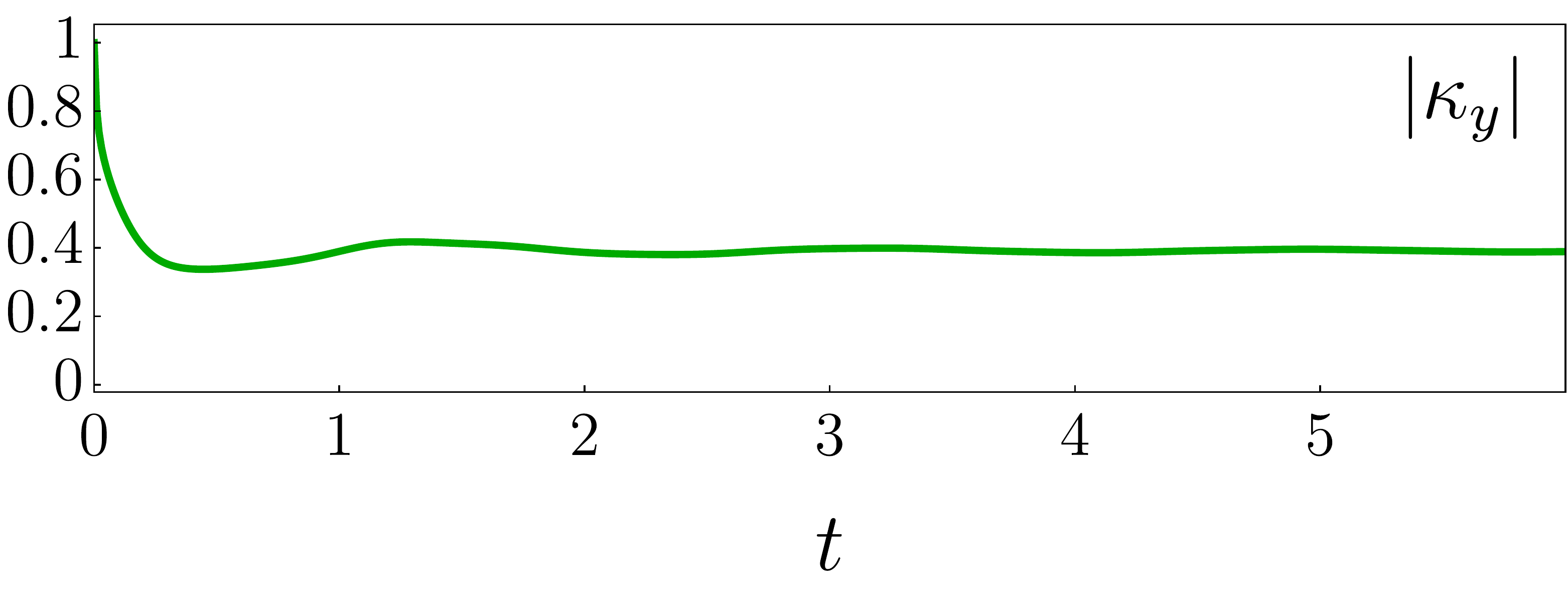}
\caption{(Color online) Coherence trapping dynamics. Magnitudes of the original decoherence function ($\kappa$) and $\mathrm{B}+$ decomposition decoherence functions ($\kappa_0, \kappa_x, \kappa_y$) as a function of time. We set $C_h=C_v=1/\sqrt{2}$.}
\label{fig7}
\end{figure}

The non-Markovian, Markovian, and coherence trapping cases are plotted respectively in Figs.~\ref{fig5}, \ref{fig6}, and \ref{fig7}. Looking at the Fig.~\ref{fig5}, one finds that $\vert \kappa(t) \vert$ first decays to zero and then it revives again. This situation displays non-Markovian features, where coherence can revive after a period of disappearance. The Markovian case however illustrates a monotonically decaying  $\vert \kappa (t) \vert$, see Fig.~\ref{fig6}. Finally in the coherence trapping case we observe that $\vert \kappa(t) \vert$ decays at first but mostly maintains its value later. 
Magnitudes of the other three decoherence 
functions, used in the $\mathrm{B+}$ decomposition, are also plotted in the corresponding figures. We observe that these decoherence functions behave similarly, in contrast to the case of non-positive map. Again, whenever $\kappa_0$, $\kappa_x$, and $\kappa_y$ are all zero, one has $\kappa(t)=0$. However since $w_x$ and $w_y$ are not equal, we get a non-zero $\kappa$, even though, $\kappa_0$, $\kappa_x$, and $\kappa_y$ seem to be identical in some regions.
%%%%%%%%%%%%%%%%%%%%%%%%%%%%%
%%%%%%%%%%%%%%%%%%%%%%%%%%%%%
%%%%%%%%%%%%%%%%%%%%%%%%%%%%%%%%%%%%%%%%%%%%%%%%%%%%%%%%%%
%%%%%%%%%%%%%%%%%%%%%%%%%%%%%%%%%%%%%%%%%%%%%%%%%%%%%%%%%%

\section{Discussion} \label{sec5}
We have studied the influence of initial correlations on open system dynamics from two perspectives corresponding to master equation descriptions and recently introduced $\mathrm{B+}$ decomposition method. By using a common two-photon dephasing scenario with local polarization-frequency interaction, our results show explicitly how initial correlations -- between the composite environments (frequencies) -- influence the decoherence rates and operator form of the master equation for the polarization state.  When the environment has a single-peak gaussian structure, the master equation contains two sets of jump operators, corresponding to sum and difference between the local interactions, and  whose weights can be controlled by changing the amount of initial environmental correlations. Here, the dephasing rates are non-negative and depend linearly on time for the considered case. Having a double-peak bivariate structure, the situation changes drastically. This opens an additional dephasing path with a non-local form for the corresponding operator, and the associated rate also has divergences. Moreover, the rates for the other two dephasing operators have distinctive functional forms.

 $\mathrm{B+}$ decomposition method, in turn, allows to study such cases where the system and environment are initially correlated, preventing the use of conventional CP-maps.  We have used this decomposition to study dephasing, when polarization and frequency of a single photon are initially correlated.
The results display in detail how the initial correlations change the dephasing contribution arising solely on initial factorized state.
Indeed, instead of having one decoherence function associated to dephasing, we have now three different decoherence functions corresponding to the elements of the 
 $\mathrm{B+}$  decomposition. Here, one of the functions arises due to initially factorized part and additional two decoherence functions include also contributions from initial polarization-frequency correlations. In general, our results shed light and help in understanding how different types of correlations influence the dephasing dynamics within the commonly used photonic framework.

%%%%%%%%%%%%%%%%%%%%%%%%%%%%%%%%%%%%%%%%%%%%%%%%%%%%%%%%%%
\section*{Acknowledgments}
Sina Hamedani Raja acknowledges finical support from Finnish Cultural Foundation. Anil Shaji acknowledges the support of the EU through the Erasmus+ program and support of the Science and Engineering Research Board, Government of India through EMR grant No. EMR/2016/007221.
%%%%%%%%%%%%%%%%%%%%%%%%%%%%%%%%%%%%%%%%%%%%%%%%%%%%%%%%%%
\section*{Appendix}
General expressions for the elements of the nonzero subspace of the decay rate matrix, denoted by $R(t)$, are
\begin{widetext}
\begin{eqnarray}
&&R_{11}(t)=-\Re{[\bold{k}_b(t)]},
\\
&&R_{12}(t)=\frac{1}{\sqrt{3}}\Big(i\Im{[\bold{k}_b(t)]}-i\Im{[\bold{k}_a(t)]}+i\Im{[\bold{\Gamma}_{ab}(t)]}+\Re{[\bold{\Gamma}_{ab}(t)]}-\Re{[\bold{k}_{a}(t)]}\Big),
\\
&&R_{13}(t)=\frac{1}{2\sqrt{6}}\Big(4i\Im{[\bold{k}_a(t)]}+2i\Im{[\bold{k}_b(t)]}-3i\Im{[\bold{k}_{ab}(t)]}-i\Im{[\bold{\Gamma}_{ab}(t)]}+4\Re{[\bold{k}_{a}(t)]}-3\Re{[\bold{k}_{ab}(t)]}-\Re{[\bold{\Gamma}_{ab}(t)]}\Big),
\\
&&R_{21}(t)=R_{12}(t)^{*},
\\
&&R_{22}(t)=\frac{1}{3}\Big(-2\Re{[\bold{k}_a(t)]}+\Re{[\bold{k}_b(t)]}-2\Re{[\bold{\Gamma}_{ab}(t)]}\Big),
\\
&&R_{23}(t)=\frac{1}{6\sqrt{2}}\Big(-3i\Im{[\bold{k}_{ab}(t)]}+6i\Im{[\bold{k}_b(t)]}+3i\Im{[\bold{\Gamma}_{ab}(t)]}-4\Re{[\bold{k}_{a}(t)]}-3\Re{[\bold{k}_{ab}(t)]}+8\Re{[\bold{k}_{b}(t)]}-\Re{[\bold{\Gamma}_{ab}(t)]}\Big),
\\
&&R_{31}(t)=R_{13}(t)^{*},
\\
&&R_{32}(t)=R_{23}(t)^{*},
\\
&&R_{33}(t)=\frac{1}{6}\Big(-2\Re{[\bold{k}_a(t)]}-3\Re{[\bold{k}_{ab}(t)]}-2\Re{[\bold{k}_b(t)]}+\Re{[\bold{\Gamma}_{ab}(t)]}\Big),
\end{eqnarray}
where we have defined $\bold{k}_i(t)=\frac{1}{\kappa_i(t)}\frac{d}{dt}\kappa_i(t)$, $\bold{\Gamma}_{ab}(t)=\frac{1}{\Lambda_{ab}(t)}\frac{d}{dt}\Lambda_{ab}(t)$ and $\bold{k}_{ab}(t)=\frac{1}{\kappa_{ab}(t)}\frac{d}{dt}\kappa_{ab}(t)$.
\end{widetext}
%%%%%%%%%%%%%%%%%%%%%%%%%%%%%%%
\bibliography{DephasingDynamics}{}

%merlin.mbs apsrev4-1.bst 2010-07-25 4.21a (PWD, AO, DPC) hacked
%Control: key (0)
%Control: author (72) initials jnrlst
%Control: editor formatted (1) identically to author
%Control: production of article title (-1) disabled
%Control: page (0) single
%Control: year (1) truncated
%Control: production of eprint (0) enabled
\begin{thebibliography}{35}%
\makeatletter
\providecommand \@ifxundefined [1]{%
 \@ifx{#1\undefined}
}%
\providecommand \@ifnum [1]{%
 \ifnum #1\expandafter \@firstoftwo
 \else \expandafter \@secondoftwo
 \fi
}%
\providecommand \@ifx [1]{%
 \ifx #1\expandafter \@firstoftwo
 \else \expandafter \@secondoftwo
 \fi
}%
\providecommand \natexlab [1]{#1}%
\providecommand \enquote  [1]{``#1''}%
\providecommand \bibnamefont  [1]{#1}%
\providecommand \bibfnamefont [1]{#1}%
\providecommand \citenamefont [1]{#1}%
\providecommand \href@noop [0]{\@secondoftwo}%
\providecommand \href [0]{\begingroup \@sanitize@url \@href}%
\providecommand \@href[1]{\@@startlink{#1}\@@href}%
\providecommand \@@href[1]{\endgroup#1\@@endlink}%
\providecommand \@sanitize@url [0]{\catcode `\\12\catcode `\$12\catcode
  `\&12\catcode `\#12\catcode `\^12\catcode `\_12\catcode `\%12\relax}%
\providecommand \@@startlink[1]{}%
\providecommand \@@endlink[0]{}%
\providecommand \url  [0]{\begingroup\@sanitize@url \@url }%
\providecommand \@url [1]{\endgroup\@href {#1}{\urlprefix }}%
\providecommand \urlprefix  [0]{URL }%
\providecommand \Eprint [0]{\href }%
\providecommand \doibase [0]{http://dx.doi.org/}%
\providecommand \selectlanguage [0]{\@gobble}%
\providecommand \bibinfo  [0]{\@secondoftwo}%
\providecommand \bibfield  [0]{\@secondoftwo}%
\providecommand \translation [1]{[#1]}%
\providecommand \BibitemOpen [0]{}%
\providecommand \bibitemStop [0]{}%
\providecommand \bibitemNoStop [0]{.\EOS\space}%
\providecommand \EOS [0]{\spacefactor3000\relax}%
\providecommand \BibitemShut  [1]{\csname bibitem#1\endcsname}%
\let\auto@bib@innerbib\@empty
%</preamble>
\bibitem [{\citenamefont {Breuer}\ and\ \citenamefont
  {Petruccione}(2002)}]{BreuerBook_2002}%
  \BibitemOpen
  \bibfield  {author} {\bibinfo {author} {\bibfnamefont {H.~P.}\ \bibnamefont
  {Breuer}}\ and\ \bibinfo {author} {\bibfnamefont {F.}~\bibnamefont
  {Petruccione}},\ }\href@noop {} {\emph {\bibinfo {title} {The theory of open
  quantum systems}}}\ (\bibinfo  {publisher} {Oxford University Press},\
  \bibinfo {address} {Great Clarendon Street},\ \bibinfo {year}
  {2002})\BibitemShut {NoStop}%
\bibitem [{\citenamefont {Huelga}\ \emph {et~al.}(2012)\citenamefont {Huelga}
  \emph {et~al.}}]{RivasBook_2012}%
  \BibitemOpen
  \bibfield  {author} {\bibinfo {author} {\bibfnamefont {S.~F.}\ \bibnamefont
  {Huelga}} \emph {et~al.},\ }\href@noop {} {\emph {\bibinfo {title} {Open
  Quantum Systems: An Introduction}}}\ (\bibinfo  {publisher} {Springer},\
  \bibinfo {year} {2012})\BibitemShut {NoStop}%
\bibitem [{\citenamefont {Rivas}\ \emph {et~al.}(2014)\citenamefont {Rivas},
  \citenamefont {Huelga},\ and\ \citenamefont {Plenio}}]{Rivas_2014}%
  \BibitemOpen
  \bibfield  {author} {\bibinfo {author} {\bibfnamefont {{\'{A}}.}~\bibnamefont
  {Rivas}}, \bibinfo {author} {\bibfnamefont {S.~F.}\ \bibnamefont {Huelga}}, \
  and\ \bibinfo {author} {\bibfnamefont {M.~B.}\ \bibnamefont {Plenio}},\
  }\href {\doibase 10.1088/0034-4885/77/9/094001} {\bibfield  {journal}
  {\bibinfo  {journal} {Reports on Progress in Physics}\ }\textbf {\bibinfo
  {volume} {77}},\ \bibinfo {pages} {094001} (\bibinfo {year}
  {2014})}\BibitemShut {NoStop}%
\bibitem [{\citenamefont {Breuer}\ \emph {et~al.}(2016)\citenamefont {Breuer},
  \citenamefont {Laine}, \citenamefont {Piilo},\ and\ \citenamefont
  {Vacchini}}]{Breuer_2016}%
  \BibitemOpen
  \bibfield  {author} {\bibinfo {author} {\bibfnamefont {H.-P.}\ \bibnamefont
  {Breuer}}, \bibinfo {author} {\bibfnamefont {E.-M.}\ \bibnamefont {Laine}},
  \bibinfo {author} {\bibfnamefont {J.}~\bibnamefont {Piilo}}, \ and\ \bibinfo
  {author} {\bibfnamefont {B.}~\bibnamefont {Vacchini}},\ }\href {\doibase
  10.1103/RevModPhys.88.021002} {\bibfield  {journal} {\bibinfo  {journal}
  {Rev. Mod. Phys.}\ }\textbf {\bibinfo {volume} {88}},\ \bibinfo {pages}
  {021002} (\bibinfo {year} {2016})}\BibitemShut {NoStop}%
\bibitem [{\citenamefont {Li}\ \emph {et~al.}(2018)\citenamefont {Li},
  \citenamefont {Hall},\ and\ \citenamefont {Wiseman}}]{Hall_2018}%
  \BibitemOpen
  \bibfield  {author} {\bibinfo {author} {\bibfnamefont {L.}~\bibnamefont
  {Li}}, \bibinfo {author} {\bibfnamefont {M.~J.}\ \bibnamefont {Hall}}, \ and\
  \bibinfo {author} {\bibfnamefont {H.~M.}\ \bibnamefont {Wiseman}},\ }\href
  {\doibase https://doi.org/10.1016/j.physrep.2018.07.001} {\bibfield
  {journal} {\bibinfo  {journal} {Physics Reports}\ }\textbf {\bibinfo {volume}
  {759}},\ \bibinfo {pages} {1 } (\bibinfo {year} {2018})},\ \bibinfo {note}
  {concepts of quantum non-Markovianity: A hierarchy}\BibitemShut {NoStop}%
\bibitem [{\citenamefont {Li}\ \emph {et~al.}(2019{\natexlab{a}})\citenamefont
  {Li}, \citenamefont {Guo},\ and\ \citenamefont {Piilo}}]{Li_2019}%
  \BibitemOpen
  \bibfield  {author} {\bibinfo {author} {\bibfnamefont {C.-F.}\ \bibnamefont
  {Li}}, \bibinfo {author} {\bibfnamefont {G.-C.}\ \bibnamefont {Guo}}, \ and\
  \bibinfo {author} {\bibfnamefont {J.}~\bibnamefont {Piilo}},\ }\href
  {\doibase 10.1209/0295-5075/127/50001} {\bibfield  {journal} {\bibinfo
  {journal} {{EPL} (Europhysics Letters)}\ }\textbf {\bibinfo {volume} {127}},\
  \bibinfo {pages} {50001} (\bibinfo {year} {2019}{\natexlab{a}})}\BibitemShut
  {NoStop}%
\bibitem [{\citenamefont {Li}\ \emph {et~al.}(2019{\natexlab{b}})\citenamefont
  {Li}, \citenamefont {Guo},\ and\ \citenamefont {Piilo}}]{Li_2019_EPL2}%
  \BibitemOpen
  \bibfield  {author} {\bibinfo {author} {\bibfnamefont {C.-F.}\ \bibnamefont
  {Li}}, \bibinfo {author} {\bibfnamefont {G.-C.}\ \bibnamefont {Guo}}, \ and\
  \bibinfo {author} {\bibfnamefont {J.}~\bibnamefont {Piilo}},\ }\href
  {\doibase 10.1209/0295-5075/128/30001} {\bibfield  {journal} {\bibinfo
  {journal} {{EPL} (Europhysics Letters)}\ }\textbf {\bibinfo {volume} {128}},\
  \bibinfo {pages} {30001} (\bibinfo {year} {2019}{\natexlab{b}})}\BibitemShut
  {NoStop}%
\bibitem [{\citenamefont {de~Vega}\ and\ \citenamefont
  {Alonso}(2017)}]{Vega_2017}%
  \BibitemOpen
  \bibfield  {author} {\bibinfo {author} {\bibfnamefont {I.}~\bibnamefont
  {de~Vega}}\ and\ \bibinfo {author} {\bibfnamefont {D.}~\bibnamefont
  {Alonso}},\ }\href {\doibase 10.1103/RevModPhys.89.015001} {\bibfield
  {journal} {\bibinfo  {journal} {Rev. Mod. Phys.}\ }\textbf {\bibinfo {volume}
  {89}},\ \bibinfo {pages} {015001} (\bibinfo {year} {2017})}\BibitemShut
  {NoStop}%
\bibitem [{\citenamefont {Laine}\ \emph
  {et~al.}(2012{\natexlab{a}})\citenamefont {Laine}, \citenamefont {Breuer},
  \citenamefont {Piilo}, \citenamefont {Li},\ and\ \citenamefont
  {Guo}}]{Laine_2012}%
  \BibitemOpen
  \bibfield  {author} {\bibinfo {author} {\bibfnamefont {E.-M.}\ \bibnamefont
  {Laine}}, \bibinfo {author} {\bibfnamefont {H.-P.}\ \bibnamefont {Breuer}},
  \bibinfo {author} {\bibfnamefont {J.}~\bibnamefont {Piilo}}, \bibinfo
  {author} {\bibfnamefont {C.-F.}\ \bibnamefont {Li}}, \ and\ \bibinfo {author}
  {\bibfnamefont {G.-C.}\ \bibnamefont {Guo}},\ }\href {\doibase
  10.1103/PhysRevLett.108.210402} {\bibfield  {journal} {\bibinfo  {journal}
  {Phys. Rev. Lett.}\ }\textbf {\bibinfo {volume} {108}},\ \bibinfo {pages}
  {210402} (\bibinfo {year} {2012}{\natexlab{a}})}\BibitemShut {NoStop}%
\bibitem [{\citenamefont {Liu}\ \emph {et~al.}(2013)\citenamefont {Liu},
  \citenamefont {Cao}, \citenamefont {Huang}, \citenamefont {Li}, \citenamefont
  {Guo}, \citenamefont {Laine}, \citenamefont {Breuer},\ and\ \citenamefont
  {Piilo}}]{Liu_2013}%
  \BibitemOpen
  \bibfield  {author} {\bibinfo {author} {\bibfnamefont {B.-H.}\ \bibnamefont
  {Liu}}, \bibinfo {author} {\bibfnamefont {D.-Y.}\ \bibnamefont {Cao}},
  \bibinfo {author} {\bibfnamefont {Y.-F.}\ \bibnamefont {Huang}}, \bibinfo
  {author} {\bibfnamefont {C.-F.}\ \bibnamefont {Li}}, \bibinfo {author}
  {\bibfnamefont {G.-C.}\ \bibnamefont {Guo}}, \bibinfo {author} {\bibfnamefont
  {E.-M.}\ \bibnamefont {Laine}}, \bibinfo {author} {\bibfnamefont {H.-P.}\
  \bibnamefont {Breuer}}, \ and\ \bibinfo {author} {\bibfnamefont
  {J.}~\bibnamefont {Piilo}},\ }\href {\doibase 10.1038/srep01781} {\bibfield
  {journal} {\bibinfo  {journal} {Scientific Reports}\ }\textbf {\bibinfo
  {volume} {3}},\ \bibinfo {pages} {1781} (\bibinfo {year} {2013})}\BibitemShut
  {NoStop}%
\bibitem [{\citenamefont {Laine}\ \emph {et~al.}(2014)\citenamefont {Laine},
  \citenamefont {Breuer},\ and\ \citenamefont {Piilo}}]{Laine_2014}%
  \BibitemOpen
  \bibfield  {author} {\bibinfo {author} {\bibfnamefont {E.-M.}\ \bibnamefont
  {Laine}}, \bibinfo {author} {\bibfnamefont {H.-P.}\ \bibnamefont {Breuer}}, \
  and\ \bibinfo {author} {\bibfnamefont {J.}~\bibnamefont {Piilo}},\ }\href
  {\doibase 10.1038/srep04620} {\bibfield  {journal} {\bibinfo  {journal}
  {Scientific Reports}\ }\textbf {\bibinfo {volume} {4}},\ \bibinfo {pages}
  {4620} (\bibinfo {year} {2014})}\BibitemShut {NoStop}%
\bibitem [{\citenamefont {Xiang}\ \emph {et~al.}(2014)\citenamefont {Xiang},
  \citenamefont {Hou}, \citenamefont {Li}, \citenamefont {Guo}, \citenamefont
  {Breuer}, \citenamefont {Laine},\ and\ \citenamefont {Piilo}}]{Xiang_2014}%
  \BibitemOpen
  \bibfield  {author} {\bibinfo {author} {\bibfnamefont {G.-Y.}\ \bibnamefont
  {Xiang}}, \bibinfo {author} {\bibfnamefont {Z.-B.}\ \bibnamefont {Hou}},
  \bibinfo {author} {\bibfnamefont {C.-F.}\ \bibnamefont {Li}}, \bibinfo
  {author} {\bibfnamefont {G.-C.}\ \bibnamefont {Guo}}, \bibinfo {author}
  {\bibfnamefont {H.-P.}\ \bibnamefont {Breuer}}, \bibinfo {author}
  {\bibfnamefont {E.-M.}\ \bibnamefont {Laine}}, \ and\ \bibinfo {author}
  {\bibfnamefont {J.}~\bibnamefont {Piilo}},\ }\href {\doibase
  10.1209/0295-5075/107/54006} {\bibfield  {journal} {\bibinfo  {journal}
  {{EPL} (Europhysics Letters)}\ }\textbf {\bibinfo {volume} {107}},\ \bibinfo
  {pages} {54006} (\bibinfo {year} {2014})}\BibitemShut {NoStop}%
\bibitem [{\citenamefont {Liu}\ \emph {et~al.}(2016)\citenamefont {Liu},
  \citenamefont {Hu}, \citenamefont {Huang}, \citenamefont {Li}, \citenamefont
  {Guo}, \citenamefont {Karlsson}, \citenamefont {Laine}, \citenamefont
  {Maniscalco}, \citenamefont {Macchiavello},\ and\ \citenamefont
  {Piilo}}]{Liu_2016}%
  \BibitemOpen
  \bibfield  {author} {\bibinfo {author} {\bibfnamefont {B.-H.}\ \bibnamefont
  {Liu}}, \bibinfo {author} {\bibfnamefont {X.-M.}\ \bibnamefont {Hu}},
  \bibinfo {author} {\bibfnamefont {Y.-F.}\ \bibnamefont {Huang}}, \bibinfo
  {author} {\bibfnamefont {C.-F.}\ \bibnamefont {Li}}, \bibinfo {author}
  {\bibfnamefont {G.-C.}\ \bibnamefont {Guo}}, \bibinfo {author} {\bibfnamefont
  {A.}~\bibnamefont {Karlsson}}, \bibinfo {author} {\bibfnamefont {E.-M.}\
  \bibnamefont {Laine}}, \bibinfo {author} {\bibfnamefont {S.}~\bibnamefont
  {Maniscalco}}, \bibinfo {author} {\bibfnamefont {C.}~\bibnamefont
  {Macchiavello}}, \ and\ \bibinfo {author} {\bibfnamefont {J.}~\bibnamefont
  {Piilo}},\ }\href {\doibase 10.1209/0295-5075/114/10005} {\bibfield
  {journal} {\bibinfo  {journal} {{EPL} (Europhysics Letters)}\ }\textbf
  {\bibinfo {volume} {114}},\ \bibinfo {pages} {10005} (\bibinfo {year}
  {2016})}\BibitemShut {NoStop}%
\bibitem [{\citenamefont {Raja}\ \emph {et~al.}(2017)\citenamefont {Raja},
  \citenamefont {Karpat}, \citenamefont {Laine}, \citenamefont {Maniscalco},
  \citenamefont {Piilo}, \citenamefont {Li},\ and\ \citenamefont
  {Guo}}]{HamedaniRaja_2017}%
  \BibitemOpen
  \bibfield  {author} {\bibinfo {author} {\bibfnamefont {S.~H.}\ \bibnamefont
  {Raja}}, \bibinfo {author} {\bibfnamefont {G.}~\bibnamefont {Karpat}},
  \bibinfo {author} {\bibfnamefont {E.-M.}\ \bibnamefont {Laine}}, \bibinfo
  {author} {\bibfnamefont {S.}~\bibnamefont {Maniscalco}}, \bibinfo {author}
  {\bibfnamefont {J.}~\bibnamefont {Piilo}}, \bibinfo {author} {\bibfnamefont
  {C.-F.}\ \bibnamefont {Li}}, \ and\ \bibinfo {author} {\bibfnamefont {G.-C.}\
  \bibnamefont {Guo}},\ }\href {\doibase 10.1103/PhysRevA.96.013844} {\bibfield
   {journal} {\bibinfo  {journal} {Phys. Rev. A}\ }\textbf {\bibinfo {volume}
  {96}},\ \bibinfo {pages} {013844} (\bibinfo {year} {2017})}\BibitemShut
  {NoStop}%
\bibitem [{\citenamefont {Pechukas}(1994)}]{pechukas94}%
  \BibitemOpen
  \bibfield  {author} {\bibinfo {author} {\bibfnamefont {P.}~\bibnamefont
  {Pechukas}},\ }\href@noop {} {\bibfield  {journal} {\bibinfo  {journal}
  {Phys. Rev. Lett.}\ }\textbf {\bibinfo {volume} {73}},\ \bibinfo {pages}
  {1060} (\bibinfo {year} {1994})}\BibitemShut {NoStop}%
\bibitem [{\citenamefont {Alicki}(1995)}]{Alicki_1995}%
  \BibitemOpen
  \bibfield  {author} {\bibinfo {author} {\bibfnamefont {R.}~\bibnamefont
  {Alicki}},\ }\href {\doibase 10.1103/PhysRevLett.75.3020} {\bibfield
  {journal} {\bibinfo  {journal} {Phys. Rev. Lett.}\ }\textbf {\bibinfo
  {volume} {75}},\ \bibinfo {pages} {3020} (\bibinfo {year}
  {1995})}\BibitemShut {NoStop}%
\bibitem [{\citenamefont {Pechukas}(1995)}]{pechukas95}%
  \BibitemOpen
  \bibfield  {author} {\bibinfo {author} {\bibfnamefont {P.}~\bibnamefont
  {Pechukas}},\ }\href@noop {} {\bibfield  {journal} {\bibinfo  {journal}
  {Phys. Rev. Lett.}\ }\textbf {\bibinfo {volume} {75}},\ \bibinfo {pages}
  {3021} (\bibinfo {year} {1995})}\BibitemShut {NoStop}%
\bibitem [{\citenamefont {Jordan}\ \emph {et~al.}(2004)\citenamefont {Jordan},
  \citenamefont {Shaji},\ and\ \citenamefont {Sudarshan}}]{jordan04}%
  \BibitemOpen
  \bibfield  {author} {\bibinfo {author} {\bibfnamefont {T.~F.}\ \bibnamefont
  {Jordan}}, \bibinfo {author} {\bibfnamefont {A.}~\bibnamefont {Shaji}}, \
  and\ \bibinfo {author} {\bibfnamefont {E.~C.~G.}\ \bibnamefont {Sudarshan}},\
  }\href@noop {} {\bibfield  {journal} {\bibinfo  {journal} {Phys. Rev A.}\
  }\textbf {\bibinfo {volume} {70}},\ \bibinfo {pages} {052110} (\bibinfo
  {year} {2004})}\BibitemShut {NoStop}%
\bibitem [{\citenamefont {Shaji}\ and\ \citenamefont
  {Sudarshan}(2005)}]{shaji05a}%
  \BibitemOpen
  \bibfield  {author} {\bibinfo {author} {\bibfnamefont {A.}~\bibnamefont
  {Shaji}}\ and\ \bibinfo {author} {\bibfnamefont {E.~C.~G.}\ \bibnamefont
  {Sudarshan}},\ }\href@noop {} {\bibfield  {journal} {\bibinfo  {journal}
  {Phys. Lett. A.}\ }\textbf {\bibinfo {volume} {341}},\ \bibinfo {pages} {48}
  (\bibinfo {year} {2005})}\BibitemShut {NoStop}%
\bibitem [{\citenamefont {Joseph}\ and\ \citenamefont
  {Shaji}(2018)}]{Linta-Shaji}%
  \BibitemOpen
  \bibfield  {author} {\bibinfo {author} {\bibfnamefont {L.}~\bibnamefont
  {Joseph}}\ and\ \bibinfo {author} {\bibfnamefont {A.}~\bibnamefont {Shaji}},\
  }\href {\doibase 10.1103/PhysRevA.97.032127} {\bibfield  {journal} {\bibinfo
  {journal} {Phys. Rev. A}\ }\textbf {\bibinfo {volume} {97}},\ \bibinfo
  {pages} {032127} (\bibinfo {year} {2018})}\BibitemShut {NoStop}%
\bibitem [{\citenamefont {Paz-Silva}\ \emph {et~al.}(2019)\citenamefont
  {Paz-Silva}, \citenamefont {Hall},\ and\ \citenamefont
  {Wiseman}}]{Wiseman_2019}%
  \BibitemOpen
  \bibfield  {author} {\bibinfo {author} {\bibfnamefont {G.~A.}\ \bibnamefont
  {Paz-Silva}}, \bibinfo {author} {\bibfnamefont {M.~J.~W.}\ \bibnamefont
  {Hall}}, \ and\ \bibinfo {author} {\bibfnamefont {H.~M.}\ \bibnamefont
  {Wiseman}},\ }\href {\doibase 10.1103/PhysRevA.100.042120} {\bibfield
  {journal} {\bibinfo  {journal} {Phys. Rev. A}\ }\textbf {\bibinfo {volume}
  {100}},\ \bibinfo {pages} {042120} (\bibinfo {year} {2019})}\BibitemShut
  {NoStop}%
\bibitem [{\citenamefont {Liu}\ \emph {et~al.}(2018)\citenamefont {Liu},
  \citenamefont {Lyyra}, \citenamefont {Sun}, \citenamefont {Liu},
  \citenamefont {Li}, \citenamefont {Guo}, \citenamefont {Maniscalco},\ and\
  \citenamefont {Piilo}}]{Lyyra_2018}%
  \BibitemOpen
  \bibfield  {author} {\bibinfo {author} {\bibfnamefont {Z.-D.}\ \bibnamefont
  {Liu}}, \bibinfo {author} {\bibfnamefont {H.}~\bibnamefont {Lyyra}}, \bibinfo
  {author} {\bibfnamefont {Y.-N.}\ \bibnamefont {Sun}}, \bibinfo {author}
  {\bibfnamefont {B.-H.}\ \bibnamefont {Liu}}, \bibinfo {author} {\bibfnamefont
  {C.-F.}\ \bibnamefont {Li}}, \bibinfo {author} {\bibfnamefont {G.-C.}\
  \bibnamefont {Guo}}, \bibinfo {author} {\bibfnamefont {S.}~\bibnamefont
  {Maniscalco}}, \ and\ \bibinfo {author} {\bibfnamefont {J.}~\bibnamefont
  {Piilo}},\ }\href@noop {} {\bibfield  {journal} {\bibinfo  {journal} {Nature
  Communications}\ }\textbf {\bibinfo {volume} {9}},\ \bibinfo {pages} {1}
  (\bibinfo {year} {2018})}\BibitemShut {NoStop}%
\bibitem [{\citenamefont {Alipour}\ \emph {et~al.}(2019)\citenamefont
  {Alipour}, \citenamefont {Rezakhani}, \citenamefont {Babu}, \citenamefont
  {Mølmer}, \citenamefont {Möttönen},\ and\ \citenamefont
  {Ala-Nissila}}]{Alipour_2019}%
  \BibitemOpen
  \bibfield  {author} {\bibinfo {author} {\bibfnamefont {S.}~\bibnamefont
  {Alipour}}, \bibinfo {author} {\bibfnamefont {A.~T.}\ \bibnamefont
  {Rezakhani}}, \bibinfo {author} {\bibfnamefont {A.~P.}\ \bibnamefont {Babu}},
  \bibinfo {author} {\bibfnamefont {K.}~\bibnamefont {Mølmer}}, \bibinfo
  {author} {\bibfnamefont {M.}~\bibnamefont {Möttönen}}, \ and\ \bibinfo
  {author} {\bibfnamefont {T.}~\bibnamefont {Ala-Nissila}},\ }\href@noop {}
  {\enquote {\bibinfo {title} {Correlation picture approach to
  open-quantum-system dynamics},}\ } (\bibinfo {year} {2019}),\ \Eprint
  {http://arxiv.org/abs/1903.03861} {arXiv:1903.03861 [quant-ph]} \BibitemShut
  {NoStop}%
\bibitem [{\citenamefont {Liu}\ \emph {et~al.}(2011)\citenamefont {Liu},
  \citenamefont {Li}, \citenamefont {Huang}, \citenamefont {Li}, \citenamefont
  {Guo}, \citenamefont {Laine}, \citenamefont {Breuer},\ and\ \citenamefont
  {Piilo}}]{ElsiNature2011}%
  \BibitemOpen
  \bibfield  {author} {\bibinfo {author} {\bibfnamefont {B.-H.}\ \bibnamefont
  {Liu}}, \bibinfo {author} {\bibfnamefont {L.}~\bibnamefont {Li}}, \bibinfo
  {author} {\bibfnamefont {Y.-F.}\ \bibnamefont {Huang}}, \bibinfo {author}
  {\bibfnamefont {C.-F.}\ \bibnamefont {Li}}, \bibinfo {author} {\bibfnamefont
  {G.-C.}\ \bibnamefont {Guo}}, \bibinfo {author} {\bibfnamefont {E.-M.}\
  \bibnamefont {Laine}}, \bibinfo {author} {\bibfnamefont {H.-P.}\ \bibnamefont
  {Breuer}}, \ and\ \bibinfo {author} {\bibfnamefont {J.}~\bibnamefont
  {Piilo}},\ }\href@noop {} {\bibfield  {journal} {\bibinfo  {journal} {Nature
  Physics}\ }\textbf {\bibinfo {volume} {7}},\ \bibinfo {pages} {931} (\bibinfo
  {year} {2011})}\BibitemShut {NoStop}%
\bibitem [{\citenamefont {Laine}\ \emph
  {et~al.}(2012{\natexlab{b}})\citenamefont {Laine}, \citenamefont {Breuer},
  \citenamefont {Piilo}, \citenamefont {Li},\ and\ \citenamefont
  {Guo}}]{ElsiPRL2012}%
  \BibitemOpen
  \bibfield  {author} {\bibinfo {author} {\bibfnamefont {E.-M.}\ \bibnamefont
  {Laine}}, \bibinfo {author} {\bibfnamefont {H.-P.}\ \bibnamefont {Breuer}},
  \bibinfo {author} {\bibfnamefont {J.}~\bibnamefont {Piilo}}, \bibinfo
  {author} {\bibfnamefont {C.-F.}\ \bibnamefont {Li}}, \ and\ \bibinfo {author}
  {\bibfnamefont {G.-C.}\ \bibnamefont {Guo}},\ }\href@noop {} {\bibfield
  {journal} {\bibinfo  {journal} {Phys. Rev. Lett.}\ }\textbf {\bibinfo
  {volume} {108}},\ \bibinfo {pages} {133} (\bibinfo {year}
  {2012}{\natexlab{b}})}\BibitemShut {NoStop}%
\bibitem [{\citenamefont {Gorini}\ \emph {et~al.}(1976)\citenamefont {Gorini},
  \citenamefont {Kossakowski},\ and\ \citenamefont
  {Sudarshan}}]{GoriniJmathPhys1976}%
  \BibitemOpen
  \bibfield  {author} {\bibinfo {author} {\bibfnamefont {V.}~\bibnamefont
  {Gorini}}, \bibinfo {author} {\bibfnamefont {A.}~\bibnamefont {Kossakowski}},
  \ and\ \bibinfo {author} {\bibfnamefont {E.~C.~G.}\ \bibnamefont
  {Sudarshan}},\ }\href@noop {} {\bibfield  {journal} {\bibinfo  {journal}
  {Journal of Mathematical Physics}\ }\textbf {\bibinfo {volume} {17}},\
  \bibinfo {pages} {821} (\bibinfo {year} {1976})}\BibitemShut {NoStop}%
\bibitem [{\citenamefont {Lindblad}(1976)}]{LindbladCommun1976}%
  \BibitemOpen
  \bibfield  {author} {\bibinfo {author} {\bibfnamefont {G.}~\bibnamefont
  {Lindblad}},\ }\href@noop {} {\bibfield  {journal} {\bibinfo  {journal}
  {Communications in Mathematical Physics}\ }\textbf {\bibinfo {volume} {48}},\
  \bibinfo {pages} {119} (\bibinfo {year} {1976})}\BibitemShut {NoStop}%
\bibitem [{\citenamefont {Guarnieri}\ \emph {et~al.}(2014)\citenamefont
  {Guarnieri}, \citenamefont {Smirne},\ and\ \citenamefont
  {Vacchini}}]{Smirne2014}%
  \BibitemOpen
  \bibfield  {author} {\bibinfo {author} {\bibfnamefont {G.}~\bibnamefont
  {Guarnieri}}, \bibinfo {author} {\bibfnamefont {A.}~\bibnamefont {Smirne}}, \
  and\ \bibinfo {author} {\bibfnamefont {B.}~\bibnamefont {Vacchini}},\
  }\href@noop {} {\bibfield  {journal} {\bibinfo  {journal} {Physical Review
  A}\ }\textbf {\bibinfo {volume} {90}},\ \bibinfo {pages} {022110} (\bibinfo
  {year} {2014})}\BibitemShut {NoStop}%
\bibitem [{\citenamefont {Smirne}\ \emph {et~al.}(2019)\citenamefont {Smirne},
  \citenamefont {Egloff}, \citenamefont {D{\'\i}az}, \citenamefont {Plenio},\
  and\ \citenamefont {Huelga}}]{Smirne2018}%
  \BibitemOpen
  \bibfield  {author} {\bibinfo {author} {\bibfnamefont {A.}~\bibnamefont
  {Smirne}}, \bibinfo {author} {\bibfnamefont {D.}~\bibnamefont {Egloff}},
  \bibinfo {author} {\bibfnamefont {M.~G.}\ \bibnamefont {D{\'\i}az}}, \bibinfo
  {author} {\bibfnamefont {M.~B.}\ \bibnamefont {Plenio}}, \ and\ \bibinfo
  {author} {\bibfnamefont {S.~F.}\ \bibnamefont {Huelga}},\ }\href@noop {}
  {\bibfield  {journal} {\bibinfo  {journal} {Quantum Science and Technology}\
  }\textbf {\bibinfo {volume} {4}},\ \bibinfo {pages} {01LT01} (\bibinfo {year}
  {2019})}\BibitemShut {NoStop}%
\bibitem [{\citenamefont {Alicki}\ and\ \citenamefont
  {Lendi}(2007)}]{Alicki1987}%
  \BibitemOpen
  \bibfield  {author} {\bibinfo {author} {\bibfnamefont {R.}~\bibnamefont
  {Alicki}}\ and\ \bibinfo {author} {\bibfnamefont {K.}~\bibnamefont {Lendi}},\
  }\href@noop {} {\emph {\bibinfo {title} {Quantum dynamical semigroups and
  applications}}},\ Vol.\ \bibinfo {volume} {717}\ (\bibinfo  {publisher}
  {Springer},\ \bibinfo {year} {2007})\BibitemShut {NoStop}%
\bibitem [{\citenamefont {Rivas}\ \emph {et~al.}(2010)\citenamefont {Rivas},
  \citenamefont {Huelga},\ and\ \citenamefont {Plenio}}]{RivasPRL2010}%
  \BibitemOpen
  \bibfield  {author} {\bibinfo {author} {\bibfnamefont {{\'A}.}~\bibnamefont
  {Rivas}}, \bibinfo {author} {\bibfnamefont {S.~F.}\ \bibnamefont {Huelga}}, \
  and\ \bibinfo {author} {\bibfnamefont {M.~B.}\ \bibnamefont {Plenio}},\
  }\href@noop {} {\bibfield  {journal} {\bibinfo  {journal} {Phys. Rev. Lett.}\
  }\textbf {\bibinfo {volume} {105}},\ \bibinfo {pages} {050403} (\bibinfo
  {year} {2010})}\BibitemShut {NoStop}%
\bibitem [{\citenamefont {Palma}\ \emph {et~al.}(1997)\citenamefont {Palma},
  \citenamefont {Suominen},\ and\ \citenamefont {Ekert}}]{PalmaPROC1996}%
  \BibitemOpen
  \bibfield  {author} {\bibinfo {author} {\bibfnamefont {G.~M.}\ \bibnamefont
  {Palma}}, \bibinfo {author} {\bibfnamefont {K.~A.}\ \bibnamefont {Suominen}},
  \ and\ \bibinfo {author} {\bibfnamefont {A.~K.}\ \bibnamefont {Ekert}},\
  }\href@noop {} {\bibfield  {journal} {\bibinfo  {journal} {Proceedings of the
  Royal Society A: Mathematical, Physical and Engineering Sciences}\ }\textbf
  {\bibinfo {volume} {452}},\ \bibinfo {pages} {567} (\bibinfo {year}
  {1997})}\BibitemShut {NoStop}%
\bibitem [{\citenamefont {Reina}\ \emph {et~al.}(2002)\citenamefont {Reina},
  \citenamefont {Quiroga},\ and\ \citenamefont {Johnson}}]{ReinaPRA2002}%
  \BibitemOpen
  \bibfield  {author} {\bibinfo {author} {\bibfnamefont {J.~H.}\ \bibnamefont
  {Reina}}, \bibinfo {author} {\bibfnamefont {L.}~\bibnamefont {Quiroga}}, \
  and\ \bibinfo {author} {\bibfnamefont {N.~F.}\ \bibnamefont {Johnson}},\
  }\href@noop {} {\bibfield  {journal} {\bibinfo  {journal} {Physical Review
  A}\ }\textbf {\bibinfo {volume} {65}},\ \bibinfo {pages} {032326} (\bibinfo
  {year} {2002})}\BibitemShut {NoStop}%
\bibitem [{\citenamefont {Cirone}\ \emph {et~al.}(2009)\citenamefont {Cirone},
  \citenamefont {De~Chiara}, \citenamefont {Palma},\ and\ \citenamefont
  {Recati}}]{CironeNEWJPhys2009}%
  \BibitemOpen
  \bibfield  {author} {\bibinfo {author} {\bibfnamefont {M.~A.}\ \bibnamefont
  {Cirone}}, \bibinfo {author} {\bibfnamefont {G.}~\bibnamefont {De~Chiara}},
  \bibinfo {author} {\bibfnamefont {G.~M.}\ \bibnamefont {Palma}}, \ and\
  \bibinfo {author} {\bibfnamefont {A.}~\bibnamefont {Recati}},\ }\href@noop {}
  {\bibfield  {journal} {\bibinfo  {journal} {New Journal of Physics}\ }\textbf
  {\bibinfo {volume} {11}},\ \bibinfo {pages} {103055} (\bibinfo {year}
  {2009})}\BibitemShut {NoStop}%
\bibitem [{\citenamefont {Addis}\ \emph {et~al.}(2013)\citenamefont {Addis},
  \citenamefont {Haikka}, \citenamefont {McEndoo}, \citenamefont
  {Macchiavello},\ and\ \citenamefont {Maniscalco}}]{AddisPRA2013}%
  \BibitemOpen
  \bibfield  {author} {\bibinfo {author} {\bibfnamefont {C.}~\bibnamefont
  {Addis}}, \bibinfo {author} {\bibfnamefont {P.}~\bibnamefont {Haikka}},
  \bibinfo {author} {\bibfnamefont {S.}~\bibnamefont {McEndoo}}, \bibinfo
  {author} {\bibfnamefont {C.}~\bibnamefont {Macchiavello}}, \ and\ \bibinfo
  {author} {\bibfnamefont {S.}~\bibnamefont {Maniscalco}},\ }\href@noop {}
  {\bibfield  {journal} {\bibinfo  {journal} {Physical Review A}\ }\textbf
  {\bibinfo {volume} {87}},\ \bibinfo {pages} {052109} (\bibinfo {year}
  {2013})}\BibitemShut {NoStop}%
\end{thebibliography}%
\end{document}